\documentclass[aps,showpacs,preprintnumbers,amsmath, amssymb]{revtex4}

\usepackage{float}
\usepackage{graphics,epsfig}
\usepackage{graphicx}
\usepackage{dcolumn}
\usepackage{bm}

\begin{document}
\baselineskip=0.8 cm
\title{{\bf Holographic p-wave superfluid in Gauss-Bonnet gravity}}

\author{Shancheng Liu$^{1,2}$, Qiyuan Pan$^{1,2}$\footnote{panqiyuan@126.com} and Jiliang Jing$^{1,2}$\footnote{jljing@hunnu.edu.cn}}
\affiliation{$^{1}$ Department of Physics, Key Laboratory of Low
Dimensional Quantum Structures and Quantum Control of Ministry of
Education, Hunan Normal University, Changsha, Hunan 410081, China}
\affiliation{$^{2}$ Synergetic Innovation Center for Quantum Effects
and Applications, Hunan Normal University, Changsha, Hunan 410081,
China}

\vspace*{0.2cm}
\begin{abstract}
\baselineskip=0.6 cm
\begin{center}
{\bf Abstract}
\end{center}

We construct the holographic p-wave superfluid in Gauss-Bonnet
gravity via a Maxwell complex vector field model and investigate the
effect of the curvature correction on the superfluid phase
transition in the probe limit. We obtain the rich phase structure
and find that the higher curvature correction hinders the condensate
of the vector field but makes it easier for the appearance of
translating point from the second-order transition to the
first-order one or for the emergence of the Cave of Winds. Moreover,
for the supercurrents versus the superfluid velocity, we observe
that our results near the critical temperature are independent of
the Gauss-Bonnet parameter and agree well with the Ginzburg-Landau
prediction.

\end{abstract}
\maketitle
\newpage
\vspace*{0.2cm}

\section{Introduction}

In recent years, the anti-de Sitter/conformal field theories
(AdS/CFT) correspondence, which can connect a strongly correlated
system in a $d$-dimensional flat spacetime with a
($d+1$)-dimensional asymptotic AdS spacetime
\cite{Maldacena,Gubser1998,Witten}, has been used to provide some
meaningful theoretical insights in order to understand the mechanism
of the high temperature superconductors from the gravitational dual
\cite{GubserPRD78}; for reviews, see Refs.
\cite{HartnollRev,HerzogRev,HorowitzRev,CaiRev} and references
therein. It was found that the so-called holographic superconductor
model, which admits black holes with scalar hair at low temperatures
(superconducting phase) but without scalar hair at high temperatures
(normal phase), turns out to be quite successful in giving the
qualitative features of the s-wave superconductivity
\cite{HartnollPRL101,HartnollJHEP12}. Introducing an $SU(2)$
Yang-Mills field into the bulk, where a gauge boson generated by one
$SU(2)$ generator is dual to the vector order parameter, the authors
of Ref. \cite{GubserPufu} presented a holographic realization of
p-wave superconductivity. Interestingly, Ref. \cite{CaiPWave-1}
constructed a new holographic p-wave superconductor model by
introducing a charged vector field into an Einstein-Maxwell theory
with a negative cosmological constant, which is a generalization of
the $SU(2)$ model with a general mass and gyromagnetic ratio
\cite{CaiPWave-2}. In Refs. \cite{DWaveChen,DWaveBenini}, the
holographic d-wave superconductivity was realized by introducing a
charged massive spin two field propagating in the bulk.

The aforementioned works on the gravitational dual models of the
superconductorlike transition focus on the vanishing spatial
components of the $U(1)$ gauge field on the AdS boundary. Since the
supercurrent in superconducting materials is very important in
condensed matter systems, it is worthwhile to construct the
corresponding holographic superfluid models by turning on the
spatial components of the gauge field according to the AdS/CFT
correspondence. Performing a deformation of the superconducting
black hole, i.e., turning on a spatial component of the gauge field
that only depends on the radial coordinate, the authors of Refs.
\cite{BasuMukherjeeShieh,HerzogKovtunSon} constructed a holographic
superfluid solution and found that the second-order superfluid phase
transition can change to the first order when the velocity of the
superfluid component increases relative to the normal component.
Arean \emph{et al.} studied the phases of the s-wave holographic
superfluids and observed the Cave of Winds phase structure where the
system first suffers a second-order transition and then a
first-order phase transition when the temperature decreases
\cite{Arean2010}. Extending the investigation to the holographic
p-wave superfluid model in the AdS black holes coupled to a Maxwell
complex vector field, Wu \emph{et al.} showed that the translating
superfluid velocity from second order to first order increases with
the increase of the mass squared of the vector field and the Cave of
Winds takes place only in the five-dimensional spacetime
\cite{PWavSuperfluidA}. Interestingly, in Ref.
\cite{LifshitzSuperfluid} the authors obtained that the Cave of
Winds appears in some range of the Lifshitz parameter even in the
four-dimensional spacetime. Other generalized investigations based
on the holographic superfluid model can be found, for example, in
Refs.
\cite{AreanJHEP2010,SonnerWithers,Zeng2011,Peng2012,KuangLiuWang,
Amado2013,Zeng2013,Amado2014,Lai2016,AriasLandea}.

As a further step along this line, it is of great interest to
generalize the investigation on the holographic superfluid model to
the Gauss-Bonnet gravity \cite{Cai-2002} and investigate the effect
of the curvature correction on the superfluid phase transition,
which will help us to understand systematically the influences of
the $1/N$ or $1/\lambda$ ($\lambda$ is the 't Hooft coupling)
corrections on the holographic dual models. It was observed that the
higher curvature correction makes the condensate of holographic
superconductors harder to form and causes the behavior of the
claimed universal ratio $\omega/T_c\approx8$ unstable
\cite{GregoryGB,Pan-WangGB2010}, which is motivated by the
application of the Mermin-Wagner theorem to the holographic
superconductors. Rich phenomena in the phase transition were also
found for the holographic superconductors in the Gauss-Bonnet
gravity where the Gauss-Bonnet parameter can play the role in
determining the order of phase transition and critical exponents in
the second order phase transition \cite{Pan-WangGB2011,JingGB2011}.
Considering the increasing interest in study of the Gauss-Bonnet
dual models \cite{Ge-Wang,Brihaye,CaiPWaveGB,Barclay2010,
Gregory2011,LiCaiZhang,KannoGB,Gangopadhyay2012,YaoJing,NieZeng,GhoraiGangopadhyay,
SheykhiSalahiMontakhab,SalahiSheykhiMontakhab,LuWuNPB2016} and
holographic p-wave models via the Maxwell complex vector field model
\cite{CaiPWave-1,CaiPWave-2}, in this work we are going to examine
the influence of the curvature correction on the holographic p-wave
superfluid model which has not been constructed as far as we know.
We will find that the Gauss-Bonnet correction affects ont only the
condensate of the vector field but also the appearance of
translating point from the second-order transition to the
first-order one or the emergence of the Cave of Winds. However, near
the critical temperature, the ratio $\left(\langle O_x
\rangle_c/{\langle O_x \rangle_\infty}\right)^2=2/3$, which is
independent of the Gauss-Bonnet correction. For simplicity and
clarity, we will concentrate on the probe limit where the
backreaction of matter fields on the spacetime metric is neglected.

The structure of this work is as follows. In Sec. II we will
construct the holographic p-wave superfluid model with the
Gauss-Bonnet corrections in the probe limit. In particular, we
derive the equations of motion and the grand potential for the
superfluid model in the Gauss-Bonnet gravity. In Sec. III we will
consider the rich phase structure of the system and investigate the
effect of the curvature correction on the superfluid phase
transition. In Sec. IV we will explore the effect of the curvature
correction on the supercurrents versus superfluid velocity. We will
conclude in the last section with our main results.

\section{Description of the holographic dual system}
In order to construct the Gauss-Bonnet holographic p-wave model of
superfluidity in the probe limit, we start with the five-dimensional
AdS black hole in the Gauss-Bonnet gravity in the form
\cite{Cai-2002}
\begin{eqnarray}\label{Gauss-BonnetBH}
ds^2=-r^{2}f(r)dt^{2}+\frac{dr^2}{r^{2}f(r)}+r^{2}(dx^{2}+dy^{2}+dz^{2}),
\end{eqnarray}
with
\begin{eqnarray}
f(r)=\frac{1}{2\alpha}\left[1-\sqrt{1-\frac{4\alpha}{L^{2}}
\left(1-\frac{r_{+}^{4}}{r^{4}}\right)}~\right],
\end{eqnarray}
where $L$ is the AdS radius, $r_{+}$ is the black hole horizon and
$\alpha$ is the Gauss-Bonnet parameter with the upper bound, i.e.,
the Chern-Simons limit $\alpha=L^{2}/4$. Note that there are the
constraints of the causality via the holographic correspondence
\cite{BrigantePRL,BuchelMyers}, we will take the range
$-7L^{2}/36\leq\alpha\leq9L^{2}/100$ for the Gauss-Bonnet parameter
in this work. In the asymptotic region ($r\rightarrow\infty$), we
have
\begin{eqnarray}
f(r)\sim\frac{1}{2\alpha}\left(1-\sqrt{1-\frac{4\alpha}{L^2}}
\right)\,,
\end{eqnarray}
so the effective asymptotic AdS scale can be defined by
\cite{Cai-2002}
\begin{eqnarray}\label{LeffAdS}
L^2_{\rm eff}=\frac{2\alpha}{1-\sqrt{1-\frac{4\alpha}{L^2}}}.
\end{eqnarray}
Obviously, the metric (\ref{Gauss-BonnetBH}) will reduce to the
Schwarzschild AdS case when $\alpha\rightarrow0$. The Hawking
temperature of the black hole, which will be interpreted as the
temperature of the CFT, can be determined by
\begin{eqnarray}
\label{Hawking temperature} T=\frac{r_{+}}{\pi L^{2}}\ .
\end{eqnarray}
For convenience, we will scale $L=1$ in the following calculation.

Considering the Maxwell complex vector field model
\cite{CaiPWave-1,CaiPWave-2}, we will build the holographic p-wave
model of superfluidity in the Gauss-Bonnet AdS black hole background
via the action
\begin{eqnarray}\label{PWaveAtion}
S=\frac{1}{16\pi G}\int
d^{5}x\sqrt{-g}\left(-\frac{1}{4}F_{\mu\nu}F^{\mu\nu}-\frac{1}{2}\rho_{\mu\nu}^{\dag}\rho^{\mu\nu}-m^2\rho_{\mu}^{\dag}\rho^{\mu}+i
q\gamma\rho_{\mu}\rho_{\nu}^{\dag}F^{\mu\nu} \right),
\end{eqnarray}
where we have defined $F_{\mu\nu}=\nabla_\mu A_{\nu}-\nabla_\nu
A_{\mu}$ and $\rho_{\mu\nu}=D_\mu\rho_\nu-D_\nu\rho_\mu$ with the
covariant derivative $D_\mu=\nabla_\mu-iqA_\mu$. $\rho_{\mu}$ is the
complex vector field with mass $m$ and charge $q$. It should be
noted that the last term, which describes the interaction between
the vector field $\rho_\mu$ and the gauge field $A_\mu$, will not
play any role since we study the case without external magnetic
field in this work.

Adopting the following ansatz for the matter fields
\begin{eqnarray}\label{Ansatz}
\rho_{\mu}dx^{\mu}=\rho_{x}(r)dx,~~A_\mu
dx^{\mu}=A_t(r)dt+A_{y}(r)dy,
\end{eqnarray}
where both a time component $A_t$ and a spatial component $A_{y}$ of
the vector potential have been introduced in order to consider the
possibility of DC supercurrent, we will obtain the equations of
motion in the probe limit
\begin{eqnarray}\label{EqrhoMotionr}
&&\rho_{x}^{\prime\prime}+\left(\frac{3}{r}+\frac{f^\prime}{f}\right)\rho_{x}^{\prime}
-\frac{1}{r^{2}f}\left(m^2+\frac{q^2A^2_y}{r^{2}}-\frac{q^2A_t^2}{r^2f}\right)\rho_{x}=0, \\
\label{EqAtMotionr}
&&A_t^{\prime\prime}+\frac{3}{r}A_t^{\prime}-\frac{2q^2\rho_{x}^2}{r^4f}A_t=0, \\
\label{EqAyMotionr}
&&A_y^{\prime\prime}+\left(\frac{3}{r}+\frac{f^\prime}{f}\right)A_y^{\prime}-\frac{2q^2\rho_{x}^2}{r^4f}A_y=0,
\end{eqnarray}
where the prime denotes the derivative with respect to $r$. From the
above equations of motion, one can easily demonstrate that the
complex vector field model is still a generalization of the $SU(2)$
Yang-Mills model in the holographic superfluid model even in the
Gauss-Bonnet gravity, which supports the argument given in
\cite{CaiPWave-2,PWavSuperfluidA}. On the other hand, Eqs.
(\ref{EqrhoMotionr}) and (\ref{EqAtMotionr}) reduce to the case
considered in \cite{LuWuNPB2016} for the holographic p-wave
conductor/superconductor phase transition in the Gauss-Bonnet
gravity, where the spatial component $A_y$ has been turned off.

We will count on the shooting method
\cite{HartnollPRL101,HartnollJHEP12} to solve numerically the
equations of motion with the appropriate boundary conditions. At the
event horizon $r=r_{+}$, the regularity condition gives
\begin{eqnarray}\label{HorizonBoundary}
A_t(r_{+})=0,~~\rho_{x}^{\prime}(r_{+})=\frac{1}{r_{+}^{2}f^\prime(r_{+})}\left[m^{2}+\frac{q^{2}A_{y}^{2}(r_{+})}{r_{+}^{2}}\right]\rho_{x}(r_{+}),
~~A_y^{\prime}(r_{+})=\frac{2q^{2}\rho_{x}^{2}(r_{+})}{r_{+}^{4}f^\prime(r_{+})}A_y(r_{+}).
\end{eqnarray}
At the asymptotic AdS boundary $r\rightarrow\infty$, we have
asymptotic behaviors
\begin{eqnarray}
\rho_{x}=\frac{\rho_{x-}}{r^{\Delta_{-}}}+\frac{\rho_{x+}}{r^{\Delta_{+}}},~~A_t=\mu-\frac{\rho}{r^{2}},
~~A_y=S_{y}-\frac{J_{y}}{r^{2}}, \label{infinity}
\end{eqnarray}
where $\Delta_\pm=1\pm\sqrt{1+m^2L^2_{\rm eff}}$ is the
characteristic exponent, $\mu$ and $S_y$ are the chemical potential
and superfluid velocity, while $\rho$ and $J_y$ are the charge
density and current in the dual field theory, respectively. Note
that $\rho_{x-}$ and $\rho_{x+}$ are interpreted as the source and
the vacuum expectation value of the vector operator $\langle
O_{x}\rangle$ in the dual field theory according to the AdS/CFT
correspondence, we will impose boundary condition $\rho_{x-}=0$
since we require that the condensate appears spontaneously.

Interestingly, from the equations of motion
(\ref{EqrhoMotionr})-(\ref{EqAyMotionr}) we can get the useful
scaling symmetries and the transformation of the relevant quantities
\begin{eqnarray}
&&r\rightarrow\lambda r\,,\hspace{0.5cm}(t, x, y,
z)\rightarrow\frac{1}{\lambda}(t, x, y,
z)\,,\hspace{0.5cm}q\rightarrow
q\,,\hspace{0.5cm}(\rho_{x},A_{t},A_{y})\rightarrow\lambda(\rho_{x},A_{t},A_{y})\,,\nonumber \\
&&(T,\mu,S_y)\rightarrow\lambda(T,\mu,S_y)\,,\hspace{0.5cm}
(\rho,J_\varphi)\rightarrow\lambda^{3}(\rho,J_\varphi)\,,\hspace{0.5cm}
\rho_{x+}\rightarrow\lambda^{1+\Delta}\rho_{x+}\,,
\label{PWSLsymmetry}
\end{eqnarray}
to build the invariant and dimensionless quantities. For simplicity,
we use the scaling symmetries to set $r_{+}=1$ and $q=1$ when
performing numerical calculations.

On the other hand, we can obtain important information about the
phase transition of the system from the behavior of the grand
potential $\Omega=-T\mathcal{S}_{os}$ of the bound state, where
$\mathcal{S}_{os}$ is the Euclidean on-shell action. As in Refs.
\cite{BasuMukherjeeShieh,HerzogKovtunSon,Arean2010,PWavSuperfluidA,LifshitzSuperfluid},
we still work in the grand canonical ensemble since we can fix the
chemical potential by considering the scaling symmetries and the
transformation (\ref{PWSLsymmetry}). From the action
(\ref{PWaveAtion}), we get
\begin{eqnarray}\label{OnShellAction}
\mathcal{S}_{os}&=&\frac{1}{16\pi G}\int dtdxdydzdr\sqrt{-g}\left[-\frac{1}{2}\nabla_\mu (A_{\nu}F^{\mu\nu})-\nabla_\mu(\rho^\dag_{\nu}\rho^{\mu\nu})+\frac{1}{2}A_\nu\nabla_\mu F^{\mu\nu}\right]\nonumber\\
&=&\frac{V_{3}}{16\pi G T}\left(-\frac{1}{2} \sqrt{-\gamma}n_r A_\nu
F^{r\nu}|_{r\rightarrow\infty}-\sqrt{-\gamma}n_r\rho^\dag_{\nu}\rho^{r\nu}|_{r\rightarrow\infty}
+\frac{1}{2}\int_{r_{+}}^\infty dr \sqrt{-g}A_\nu\nabla_\mu F^{\mu\nu}\right)\nonumber\\
&=& \frac{V_{3}}{16\pi G T}\left[\mu\rho-\frac{S_y J_y}{L^2_{\rm
eff}}+\int_{r_+}^\infty dr
\frac{\rho_{x}^2}{r}\left(A_y^2-\frac{A^2_{t}}{f}\right)\right],
\end{eqnarray}
where we have used the integration $\int dt dxdydz=V_{3}/T$. It
should be noted that we do not need to introduce the Gibbons-Hawking
boundary term for the well-defined Dirichlet variational problem and
the counterterms for the divergent terms in the on-shell action
since we neglect the backreaction of matter fields on the spacetime
metric and impose the source-free boundary condition
\cite{PWavSuperfluidA,LifshitzSuperfluid}. Ignoring the prefactor
$16\pi G$ for simplicity, we can express the grand potential in the
superfluid phase as
\begin{eqnarray}\label{GrandPotentialSuperfluid}
\frac{\Omega_{S}}{V_{3}}=-\frac{T\mathcal{S}_{os}}{V_{3}}
=-\mu\rho+\frac{S_y J_y}{L^2_{\rm eff}}+\int_{r_+}^\infty dr
\frac{\rho_{x}^2}{r}\left(\frac{A^2_{t}}{f}-A_y^2\right).
 \end{eqnarray}
Considering that $\rho_{x}=0$ in the normal phase, we can easily
obtain the grand potential in this case
\begin{eqnarray}\label{GrandPotentialNormal}
\frac{\Omega_{N}}{V_{3}}=-\mu^{2}.
 \end{eqnarray}

\section{Condensates of the vector field}

In this section, we will numerically solve the system of coupled
differential equations (\ref{EqrhoMotionr})-(\ref{EqAyMotionr}) and
obtain the condensate $\langle O_{x}\rangle$ as a function of the
temperature and the superfluid velocity. In order to determine the
critical temperature and which phase is more thermodynamically
favored, we also calculate the grand potential $\Omega$ of the bound
state.

\begin{figure}[ht]
\includegraphics[scale=0.6]{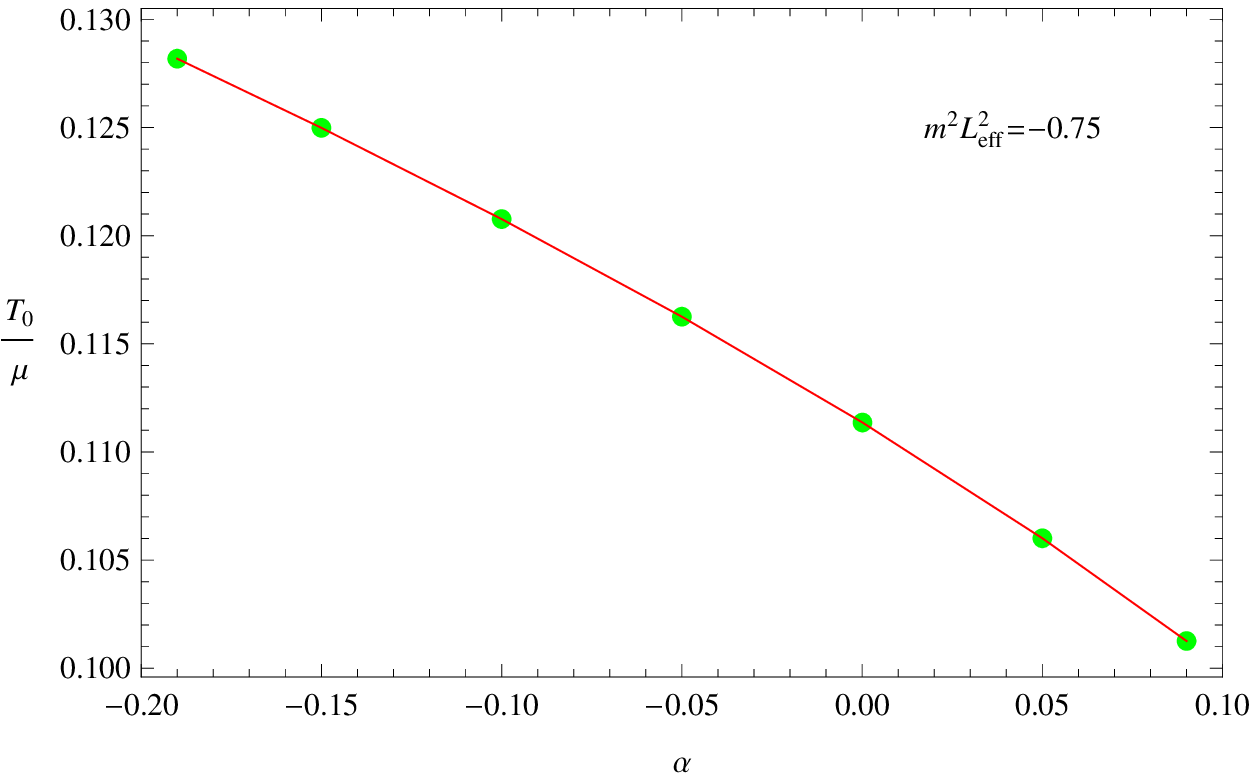}\hspace{0.2cm}%
\includegraphics[scale=0.6]{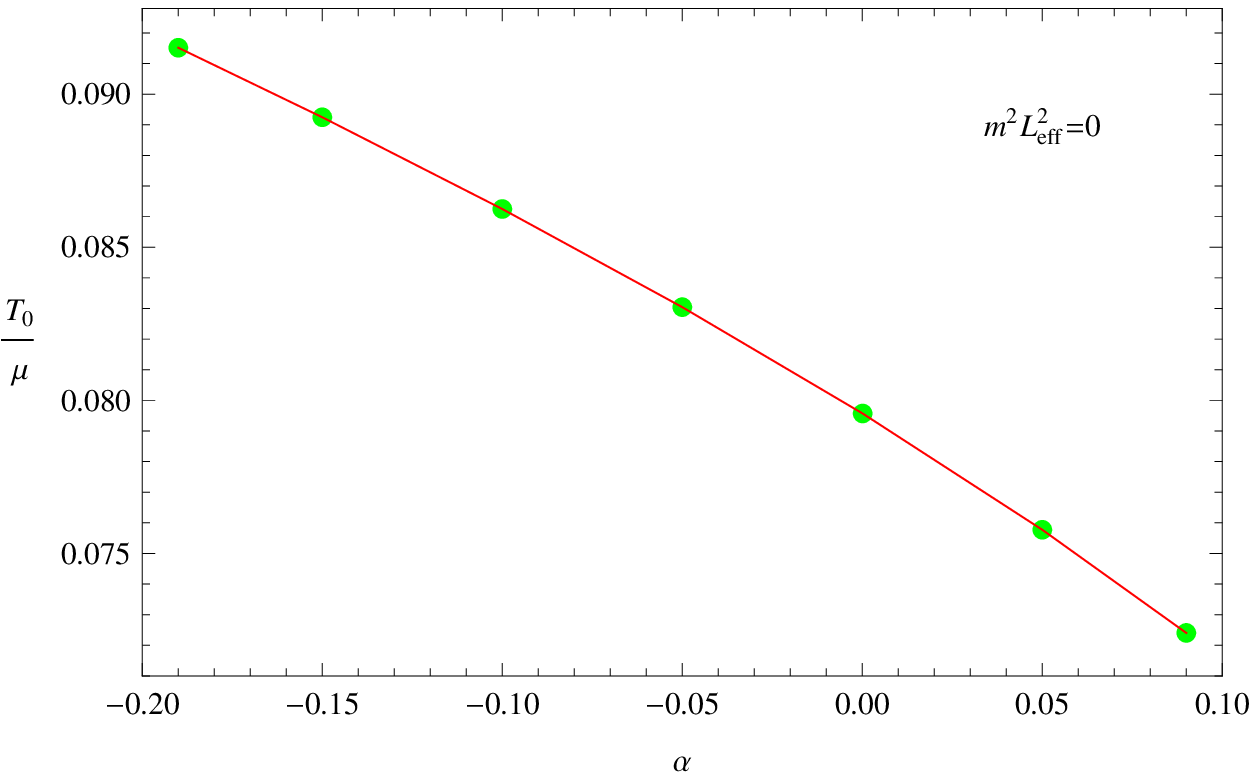}\\ \vspace{0.0cm}
\includegraphics[scale=0.6]{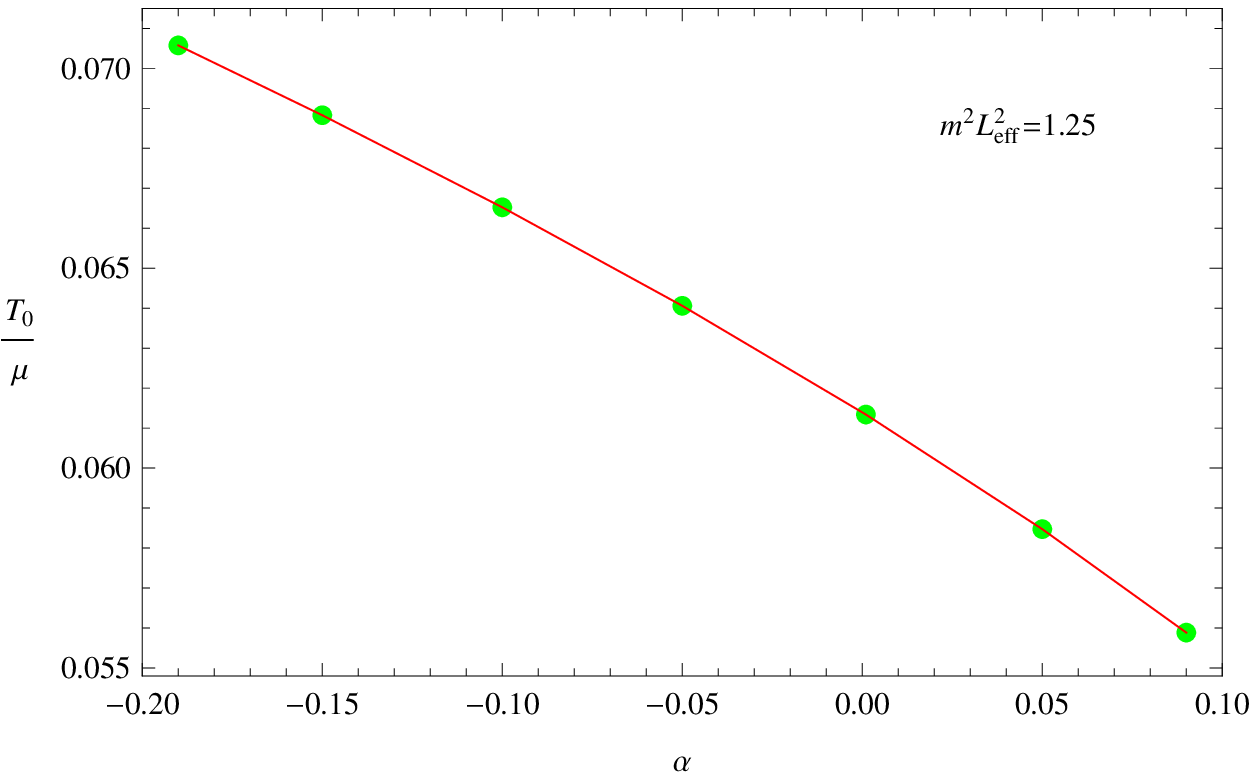}\hspace{0.2cm}%
\includegraphics[scale=0.6]{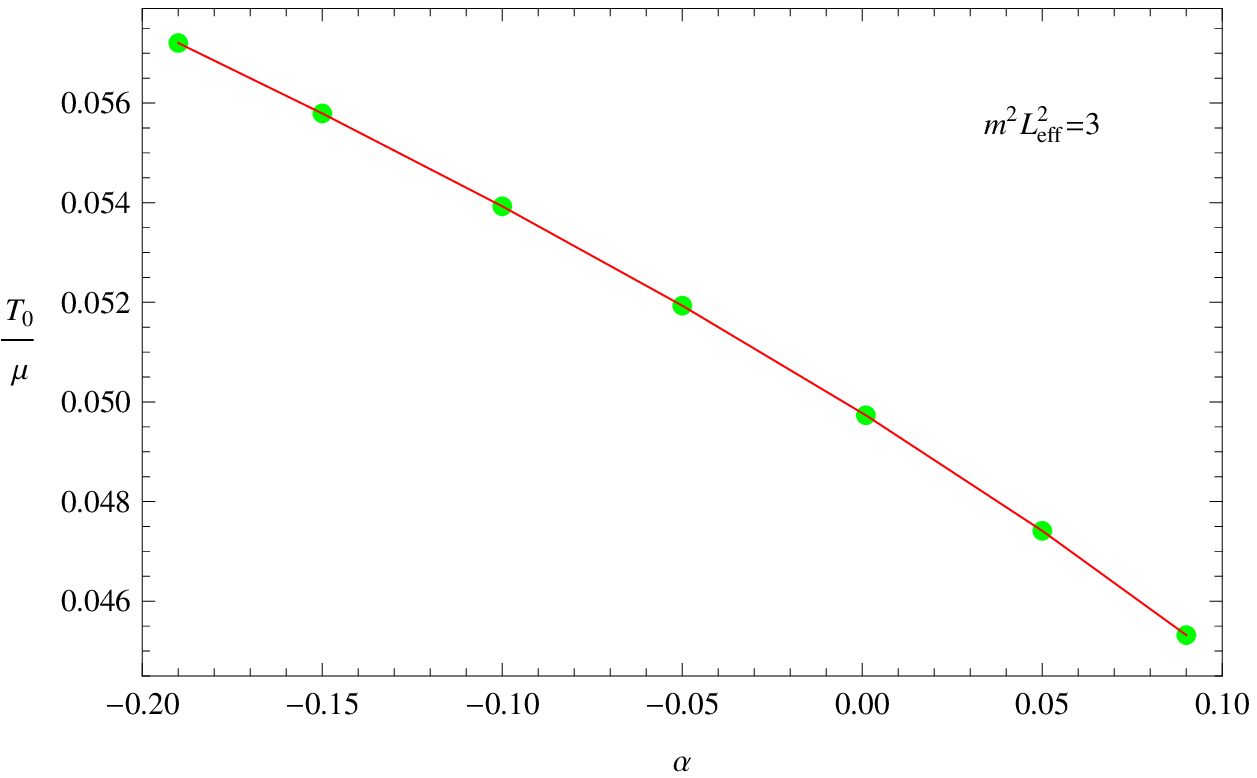}\\ \vspace{0.0cm}
\caption{\label{CriTempTc} (Color online) The critical temperature
$T_{0}$ without the superfluid velocity as a function of the
Gauss-Bonnet parameter $\alpha$ for the fixed mass of the vector
field $m^{2}L^{2}_{eff}=-3/4$, $0$, $5/4$ and $3$.}
\end{figure}

Before studying the condensed phase with nonzero $A_{y}$, we first
review the results of the conductor/superconductor phase transition
with $A_{y}=0$ in \cite{LuWuNPB2016}. It is clearly shown that, in
the absence of the superfluid velocity, the holographic
conductor/superconductor phase transition is always the second-order
one \cite{LuWuNPB2016}. In Fig. \ref{CriTempTc}, we exhibit the
critical temperature $T_{c}$ as a function of the Gauss-Bonnet
parameter $\alpha$ for the fixed mass of the vector field
$m^{2}L^{2}_{eff}=-3/4$, $0$, $5/4$ and $3$. Obviously, the critical
temperature decreases as the Gauss-Bonnet parameter increases, which
indicates that the increasing high curvature correction hinders the
conductor/superconductor phase transition. Interestingly, this
conclusion is independent of the mass of the vector field. Our
results agree well with the findings in the Gauss-Bonnet gravity for
the p-wave condensates in the Maxwell complex vector field model
\cite{LuWuNPB2016} and Yang-Mills theory \cite{CaiPWaveGB}, and the
s-wave condensates \cite{GregoryGB,Pan-WangGB2010}.

Now we are in a position to study the effects of the Gauss-Bonnet
parameter on the phase transition with superfluid velocity. In Ref.
\cite{Arean2010}, the authors found that the phase structure of the
s-wave superfluid model in the $5$D AdS black hole depends on the
range of the scalar mass, i.e., for small mass beyond the BF bound,
the phase transition changes from the second order to the first
order when the superfluid velocity increases to the translating
value; for the intermediate mass scale, the Cave of Winds appears;
and for sufficiently high mass, the phase transition is always of
the second order, no matter how high the superfluid velocity. The
holographic p-wave superfluid models in the $5$D AdS black hole
\cite{PWavSuperfluidA} and $4$D Lifshitz black hole
\cite{LifshitzSuperfluid} coupled to a Maxwell complex vector field
share similar features for the condensates of the vector field.
Thus, we will present our results in an appropriate range of the
vector field mass.

\begin{figure}[ht]
\includegraphics[scale=0.425]{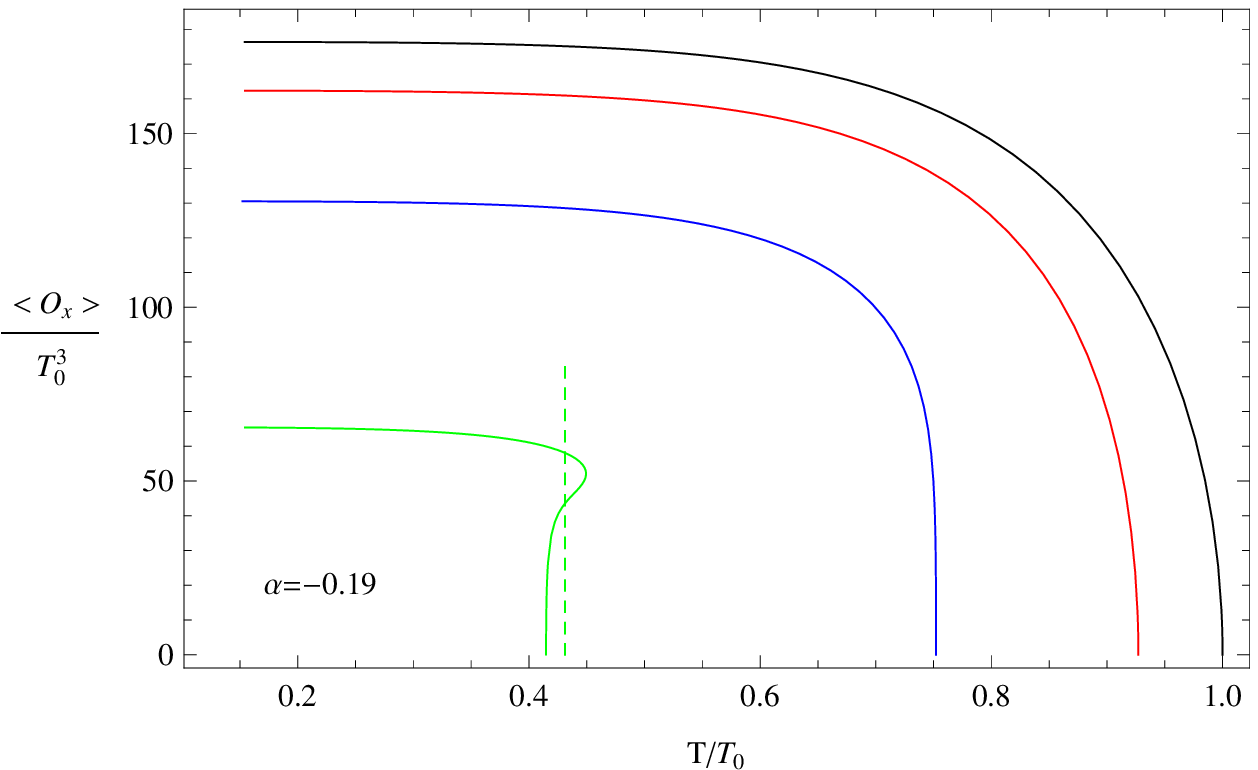}\hspace{0.2cm}%
\includegraphics[scale=0.425]{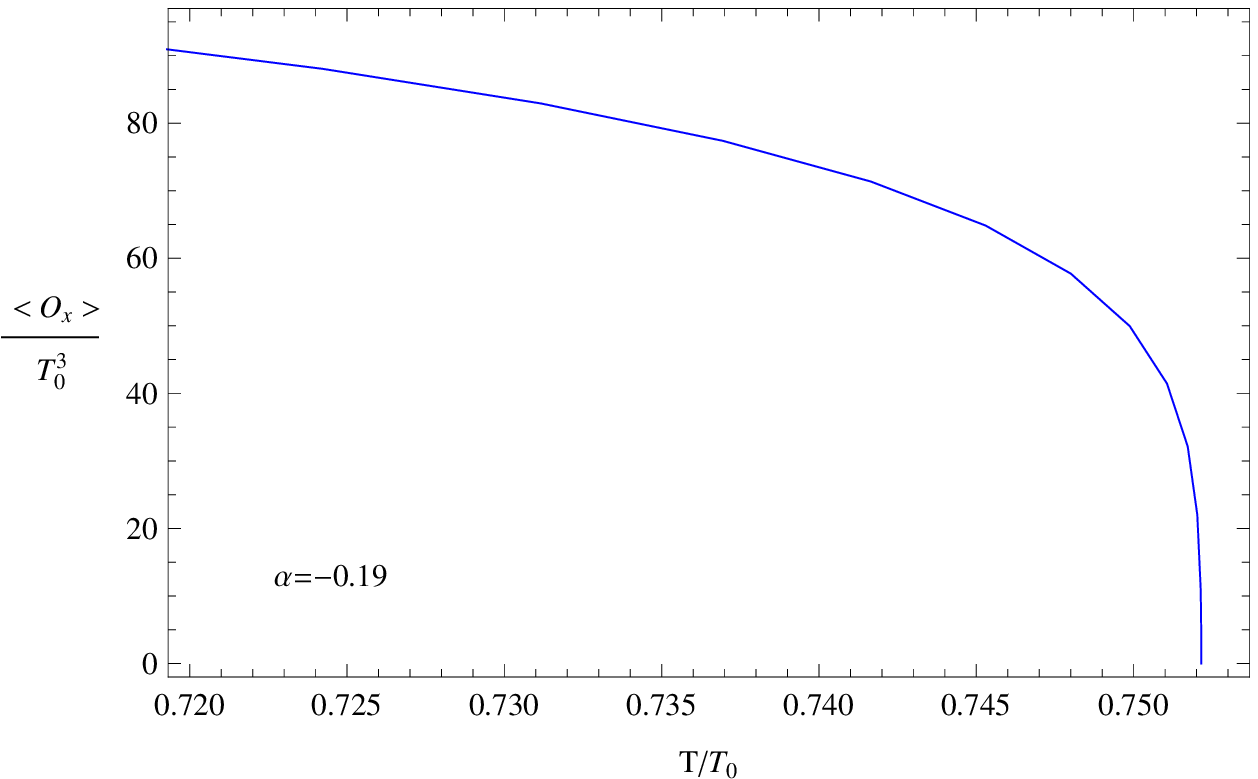}\hspace{0.2cm}%
\includegraphics[scale=0.425]{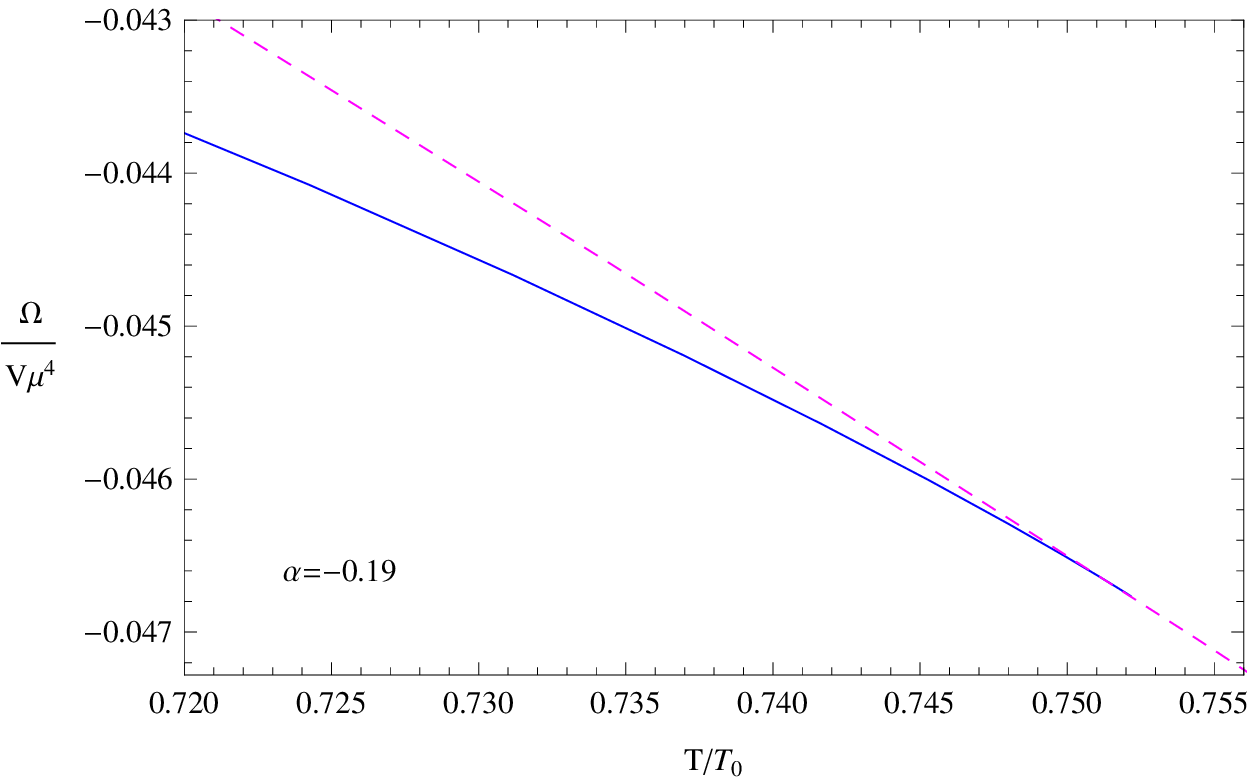}\\ \vspace{0.0cm}
\includegraphics[scale=0.425]{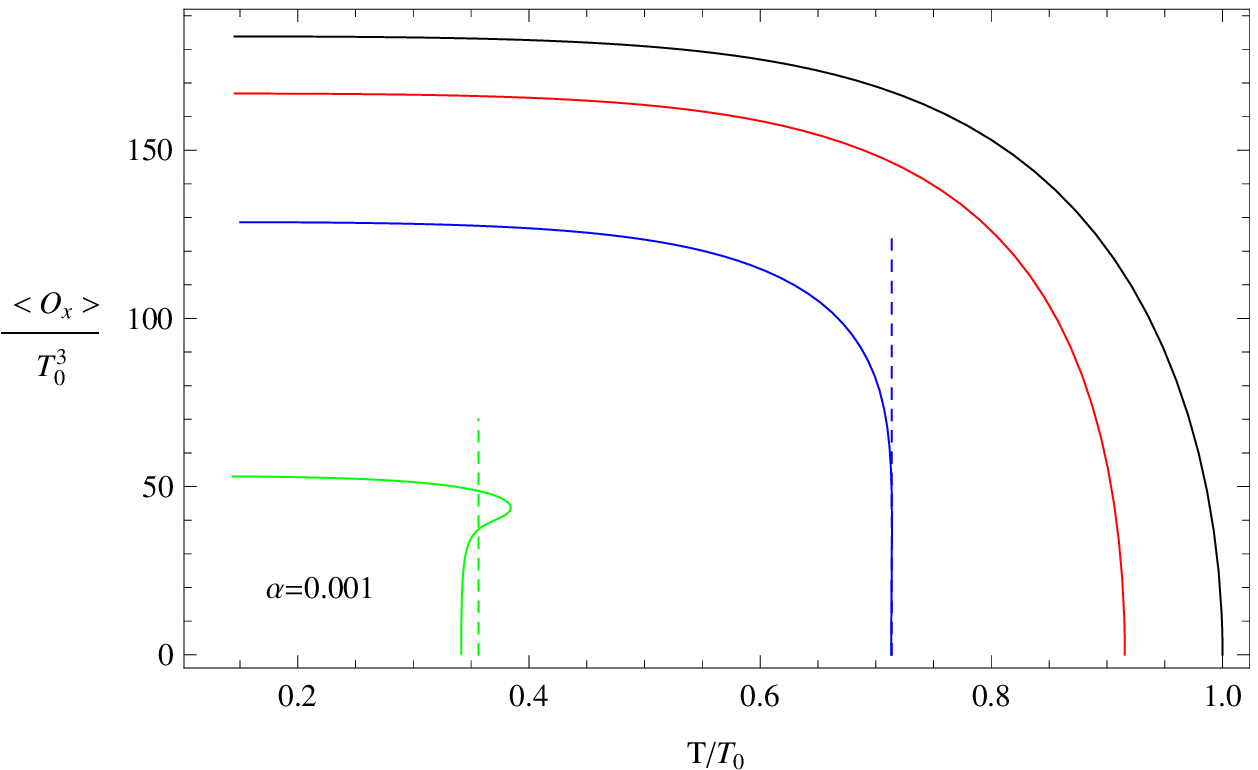}\hspace{0.2cm}%
\includegraphics[scale=0.425]{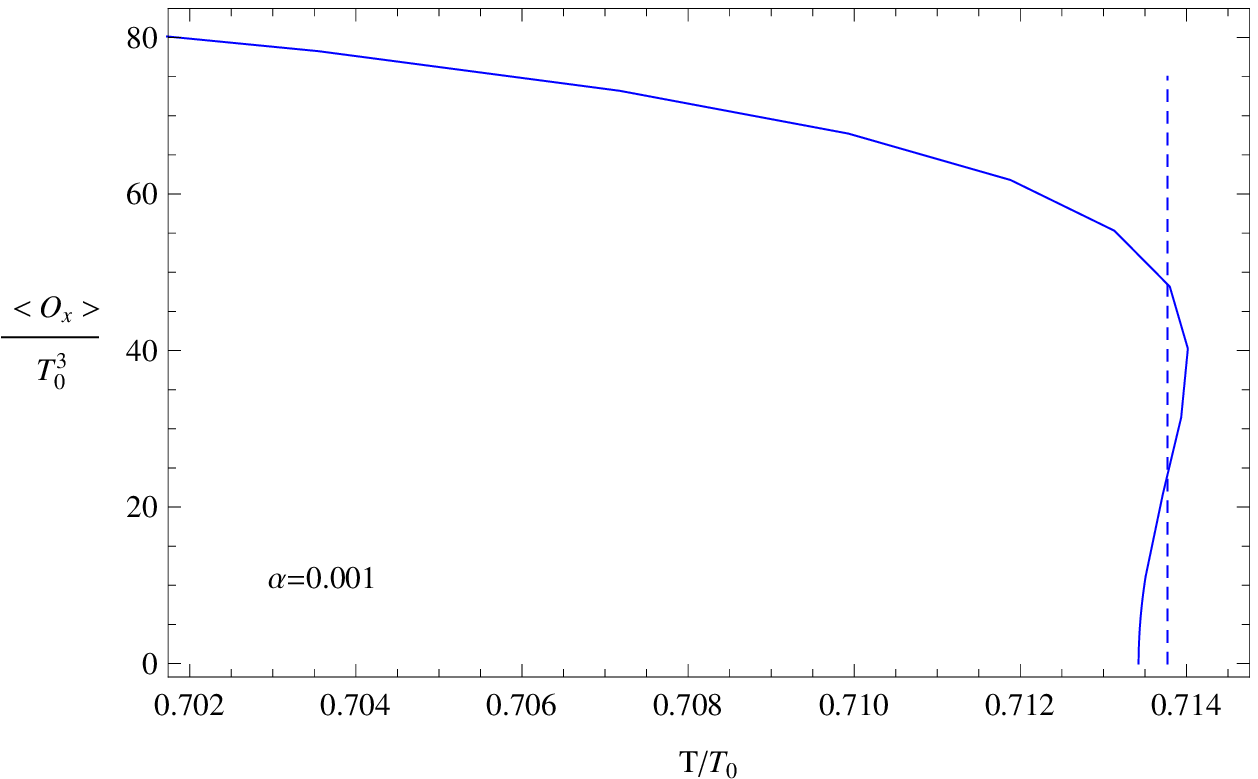}\hspace{0.2cm}%
\includegraphics[scale=0.425]{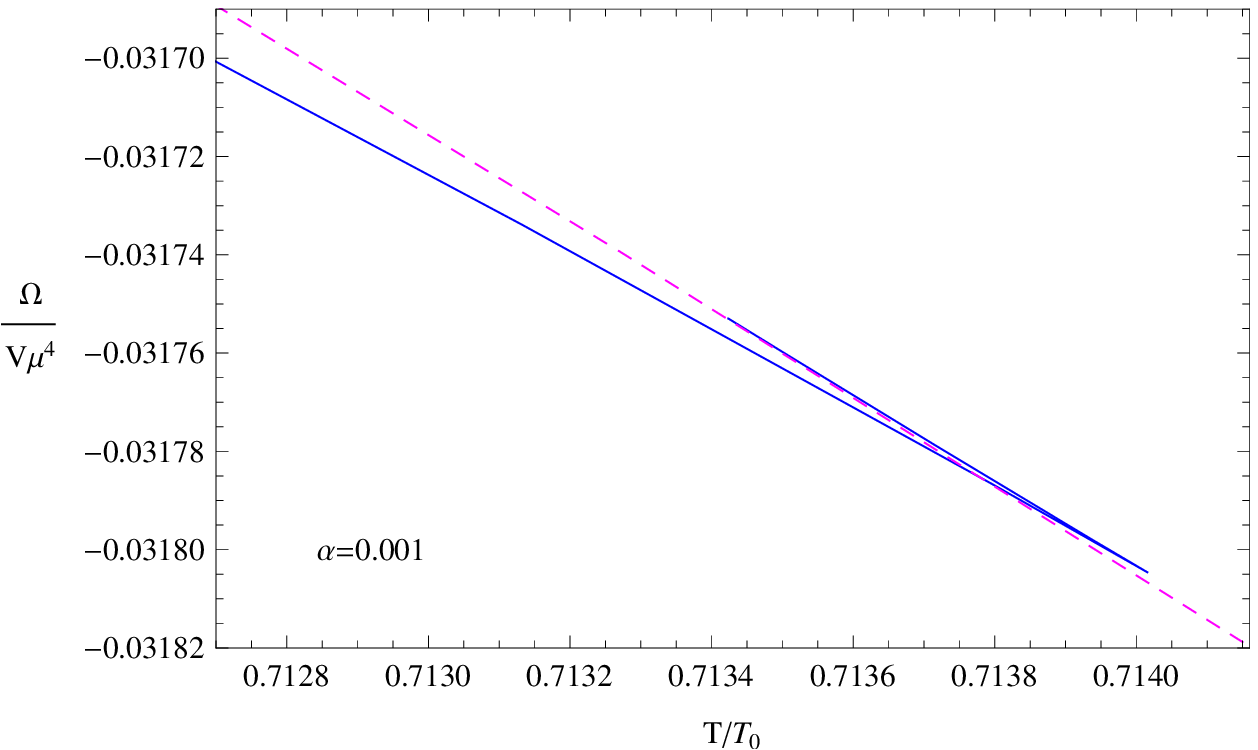}\\ \vspace{0.0cm}
\includegraphics[scale=0.425]{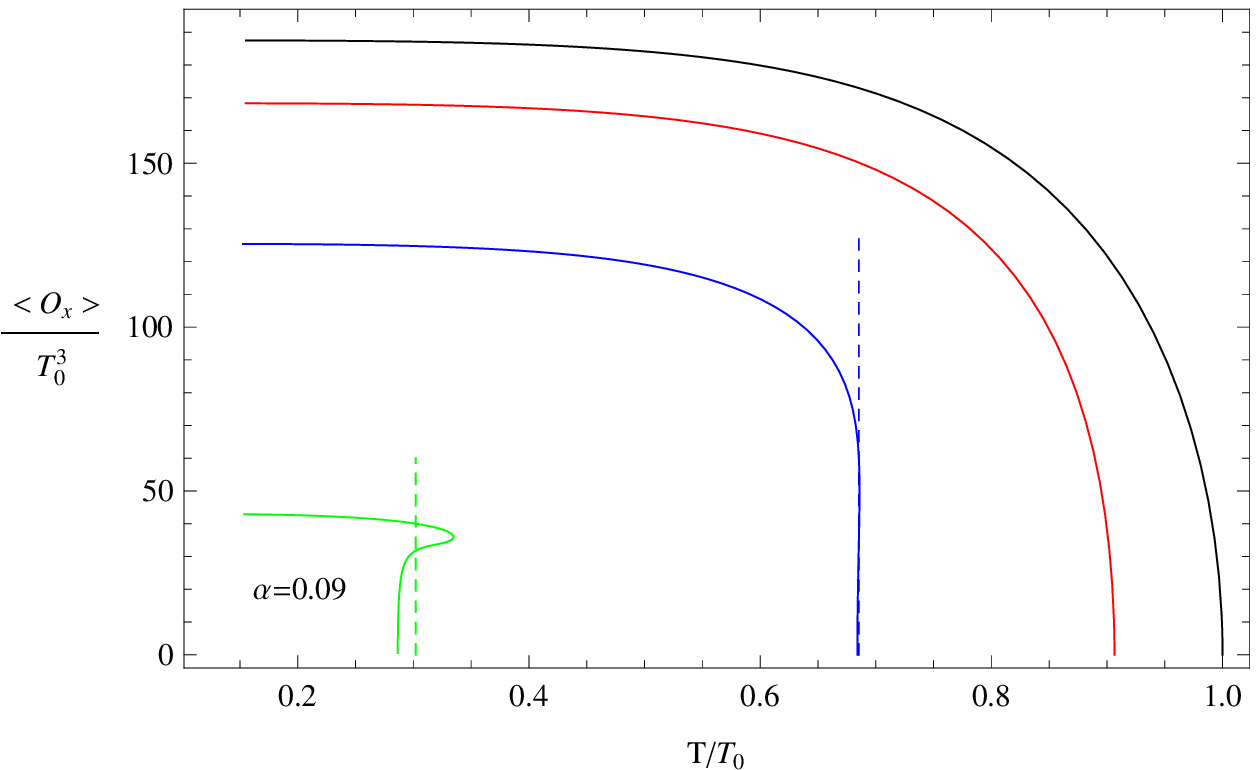}\hspace{0.2cm}%
\includegraphics[scale=0.425]{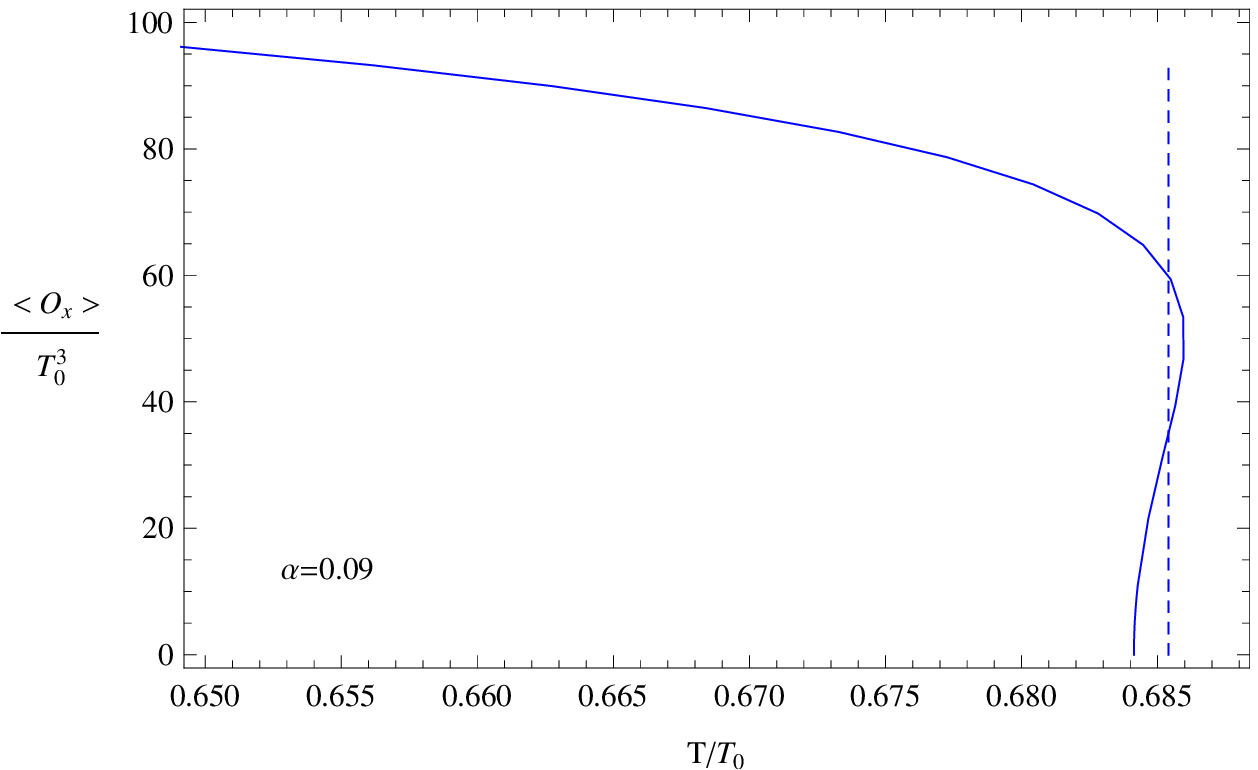}\hspace{0.2cm}%
\includegraphics[scale=0.425]{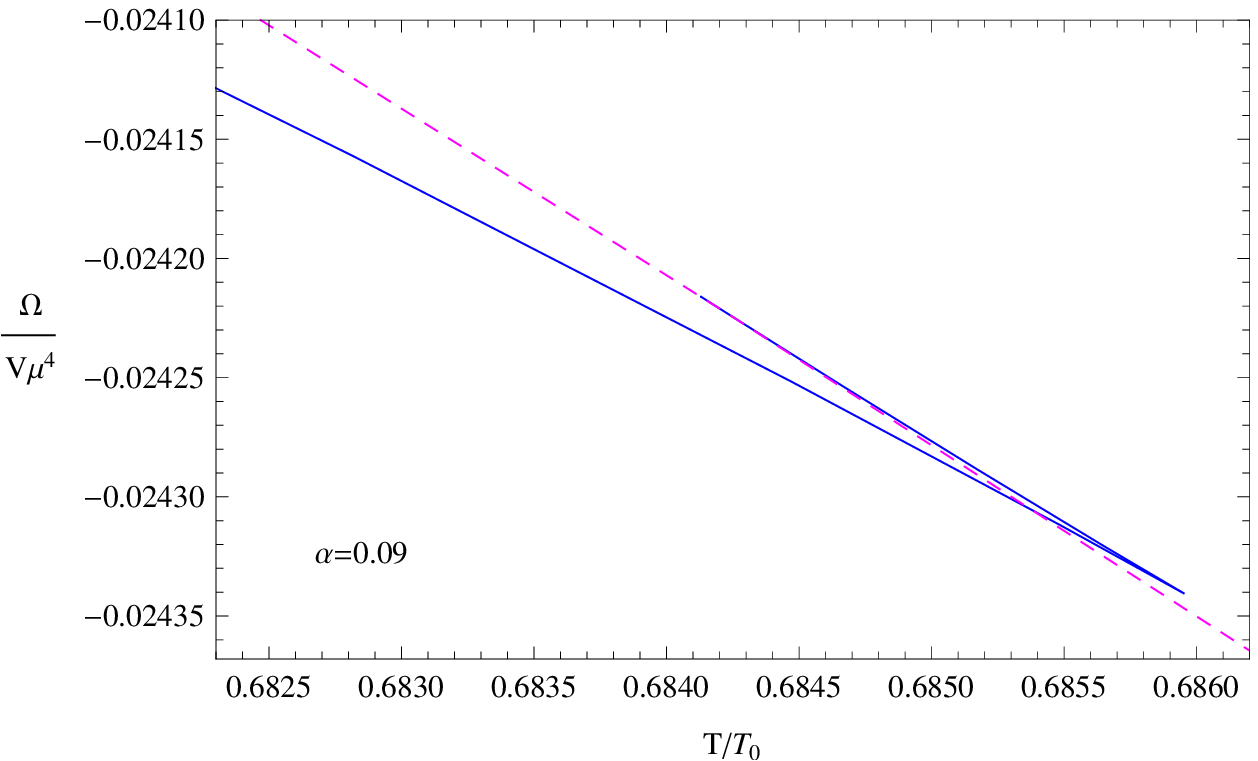}\\ \vspace{0.0cm}
\caption{\label{CondensateM0} (Color online) The condensate and the
grand potential as a function of the temperature with the fixed mass
of the vector field $m^{2}L^{2}_{eff}=0$ for different values of the
Gauss-Bonnet parameter $\alpha$. For the left three panels, the four
lines in each panel from top to bottom correspond to increasing
superfluid velocity, i.e., $\frac{S_{y}}{\mu}=0$ (black), $0.25$
(red), $0.46$ (blue) and $0.75$ (green) respectively. For the middle
three panels, the line in each panel corresponds to the superfluid
velocity $\frac{S_{y}}{\mu}=0.46$ and a vertical line represents the
critical temperature of the first-order phase transition. For the
right three panels, the two lines in each panel correspond to the
superfluid velocity $\frac{S_{y}}{\mu}=0.46$ (blue solid) and the
normal phase (magenta dotted) respectively.}
\end{figure}

\begin{figure}[ht]
\includegraphics[scale=0.65]{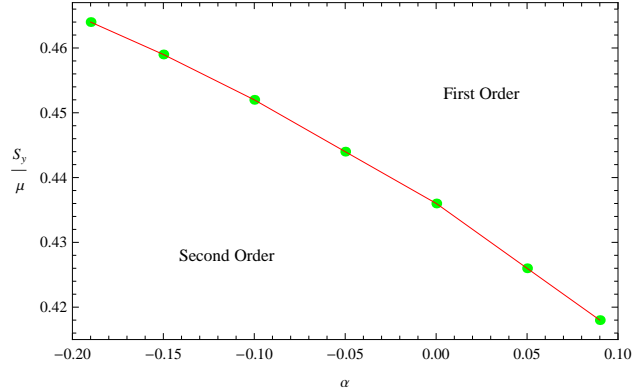} \vspace{0.0cm}
\caption{\label{TransSuperVelocity} (Color online) The translating
superfluid velocity $S_{y}/\mu$ from the second to the first order
as a function of the Gauss-Bonnet parameter $\alpha$ with the fixed
mass of the vector field $m^{2}L^{2}_{eff}=0$.}
\end{figure}

As an example of the small mass scale, we give in Fig.
\ref{CondensateM0} the condensate and the corresponding grand
potential as a function of the temperature with $m^{2}L^{2}_{eff}=0$
for different values of $\alpha$. In the case of vanishing or small
superfluid velocity, for example $\frac{S_{y}}{\mu}=0$ and $0.25$
with $\alpha=0.001$, the second-order phase transition occurs as the
temperature is lowered below a critical value, where the critical
temperature $T_{0}$ with $S_{y}=0$ is given in Fig. \ref{CriTempTc}.
The second-order superfluid phase transition will change to the
first order when superfluid velocity increases, for example
$\frac{S_{y}}{\mu}=0.46$ and $0.75$ with $\alpha=0.001$, which is
the result we can clearly observe from the grand potential with the
typical swallowtails in the bottom two panels of the rightmost
column in Fig. \ref{CondensateM0}. Obviously, there exists a
translating value of $\frac{S_{y}}{\mu}$ beyond which the phase
transition changes from the second order to the first order. In Fig.
\ref{TransSuperVelocity}, we plot the translating superfluid
velocity $S_{y}/\mu$ from the second to the first order as a
function of the Gauss-Bonnet parameter $\alpha$ with
$m^{2}L^{2}_{eff}=0$ and find that it almost decreases monotonously
with $\alpha$, which shows that the higher curvature correction
makes it easier for the emergence of the translating point. On the
other hand, we mark the locations of the critical temperature with a
vertical dotted line in the same color as the condensate curve for
the first-order phase transition in Fig. \ref{CondensateM0}. It is
found that the critical temperature decreases with the increase of
the Gauss-Bonnet parameter, which indicates that the higher
curvature correction hinders the condensate of the vector field even
in the case of the first-order phase transition.

\begin{figure}[ht]
\includegraphics[scale=0.425]{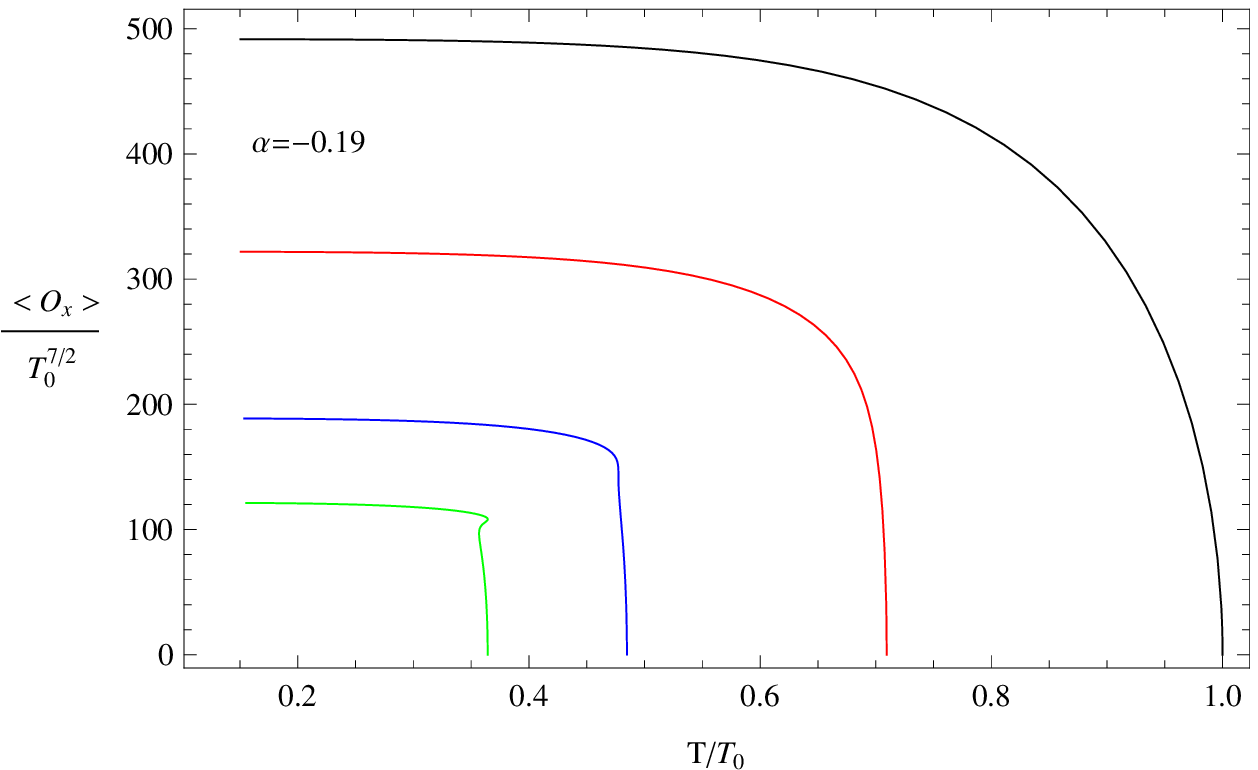}\hspace{0.2cm}%
\includegraphics[scale=0.425]{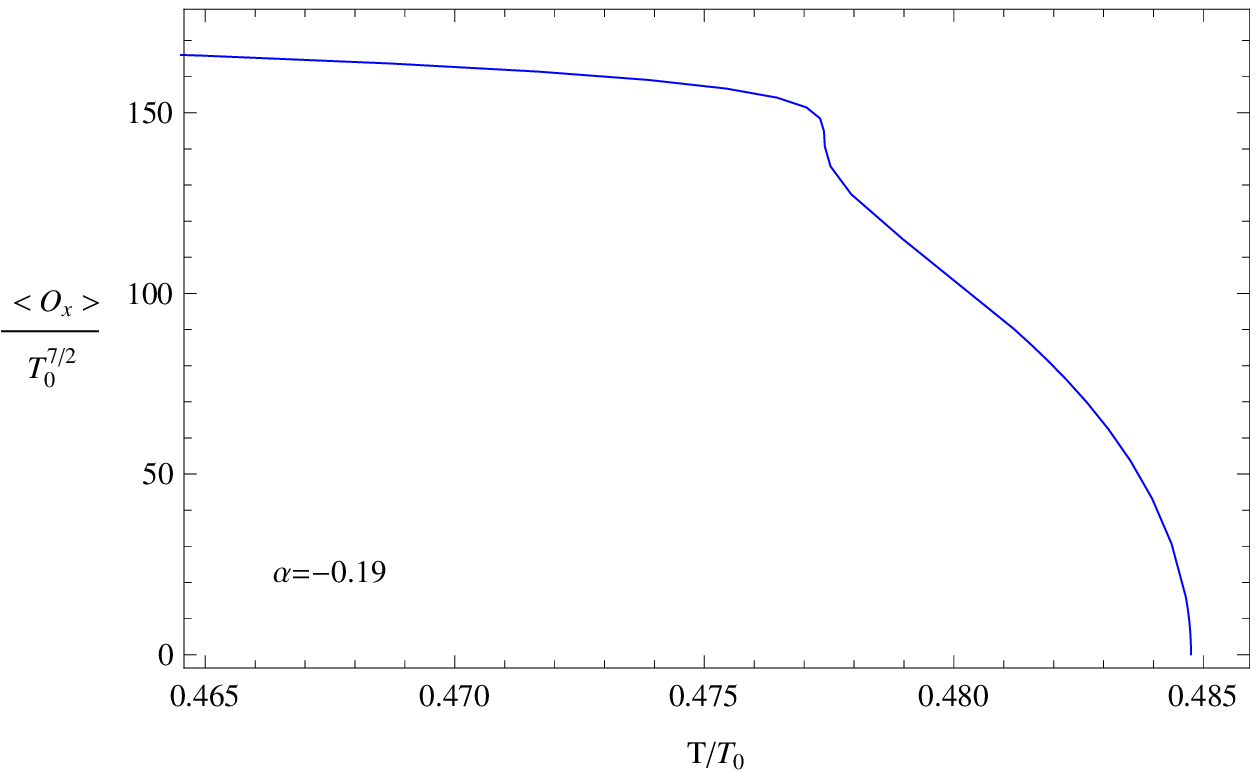}\hspace{0.2cm}%
\includegraphics[scale=0.425]{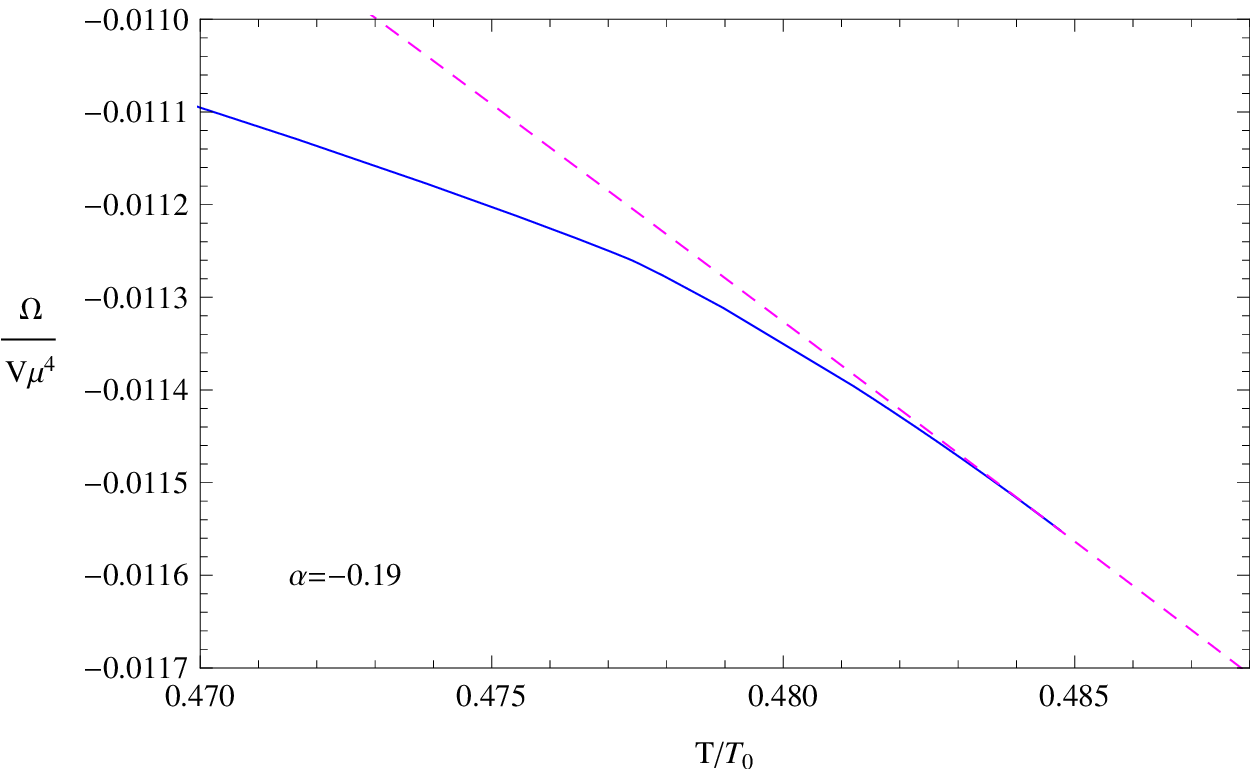}\\ \vspace{0.0cm}
\includegraphics[scale=0.425]{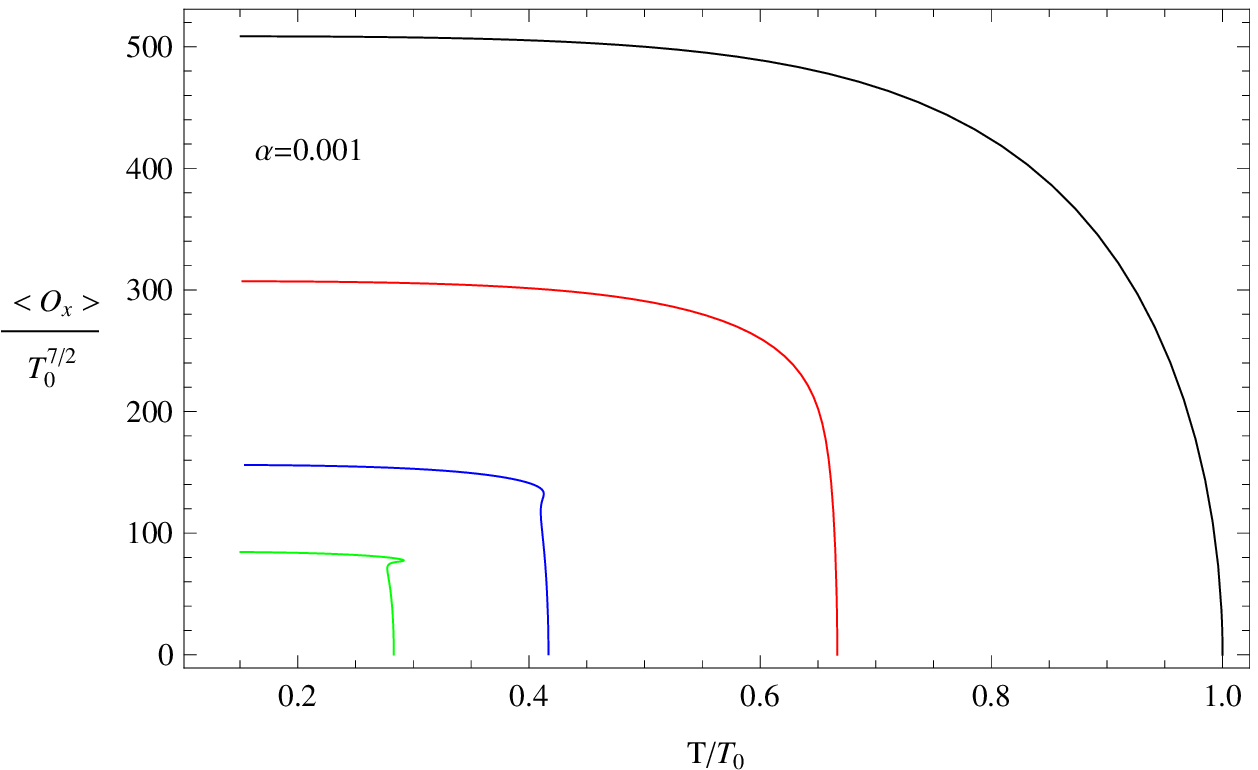}\hspace{0.2cm}%
\includegraphics[scale=0.425]{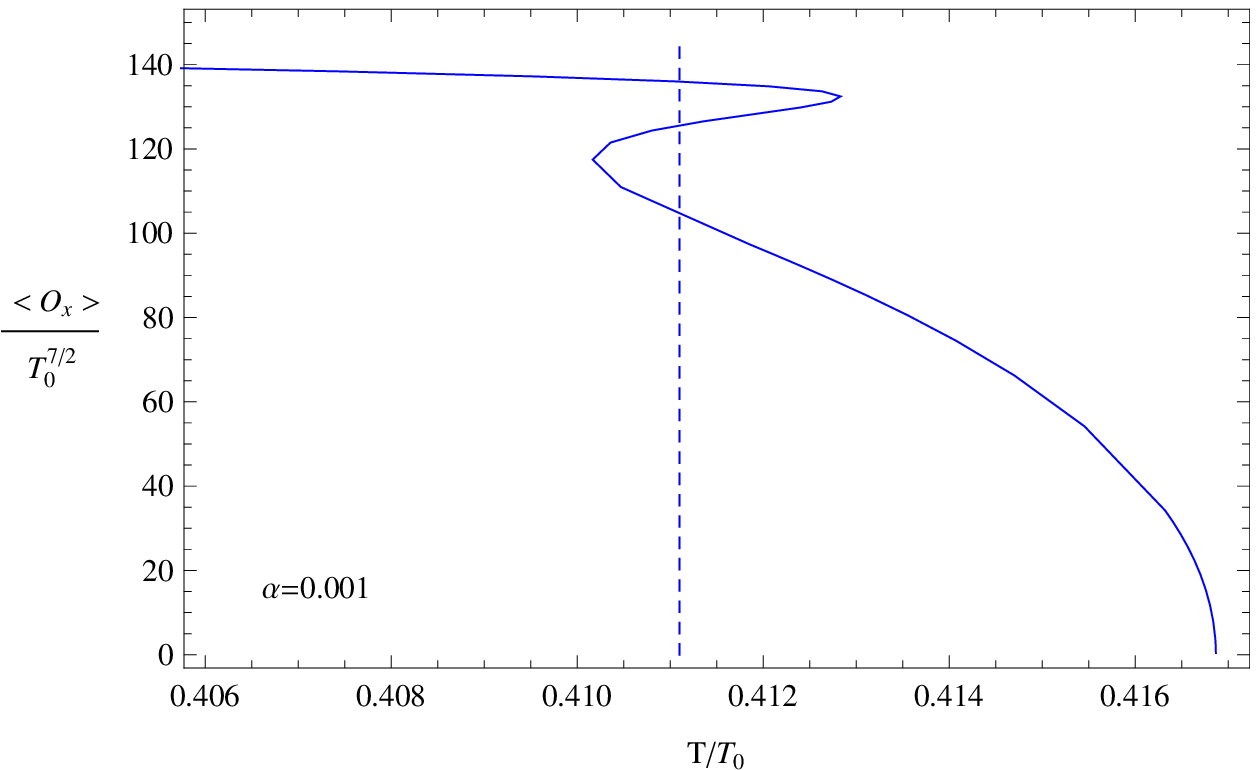}\hspace{0.2cm}%
\includegraphics[scale=0.425]{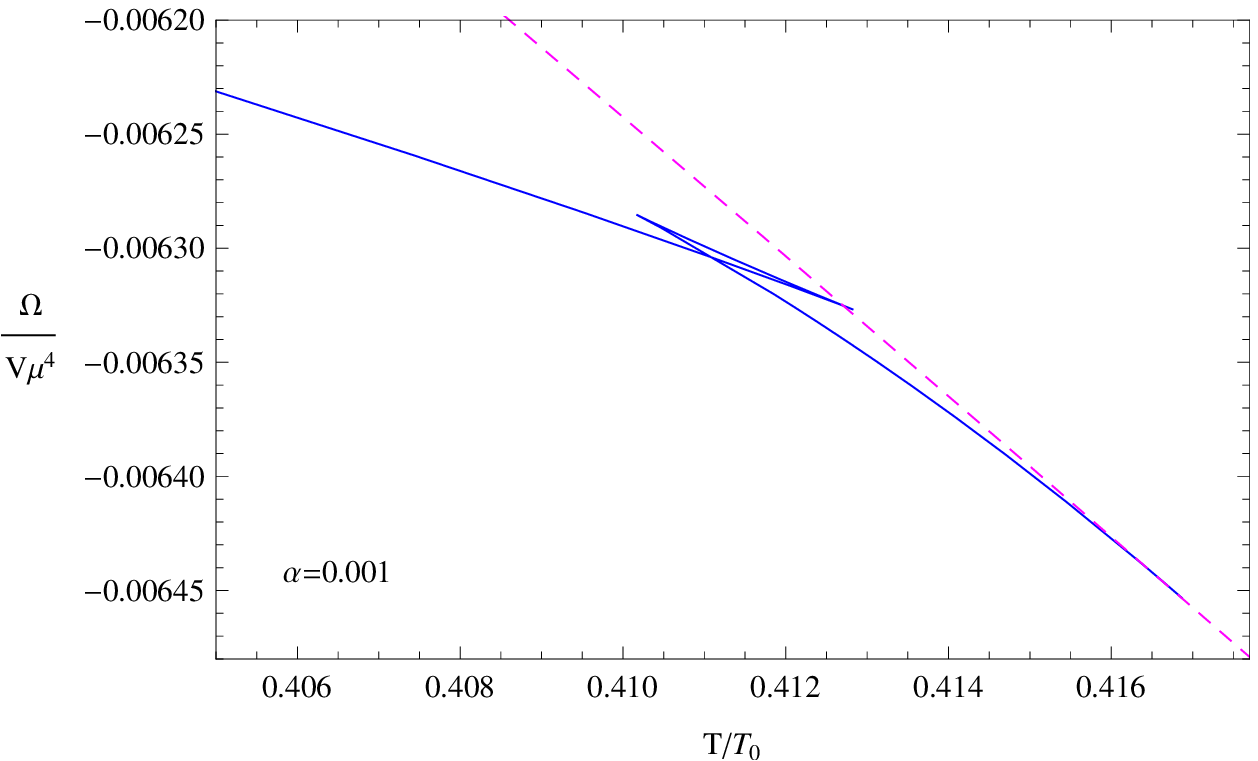}\\ \vspace{0.0cm}
\includegraphics[scale=0.425]{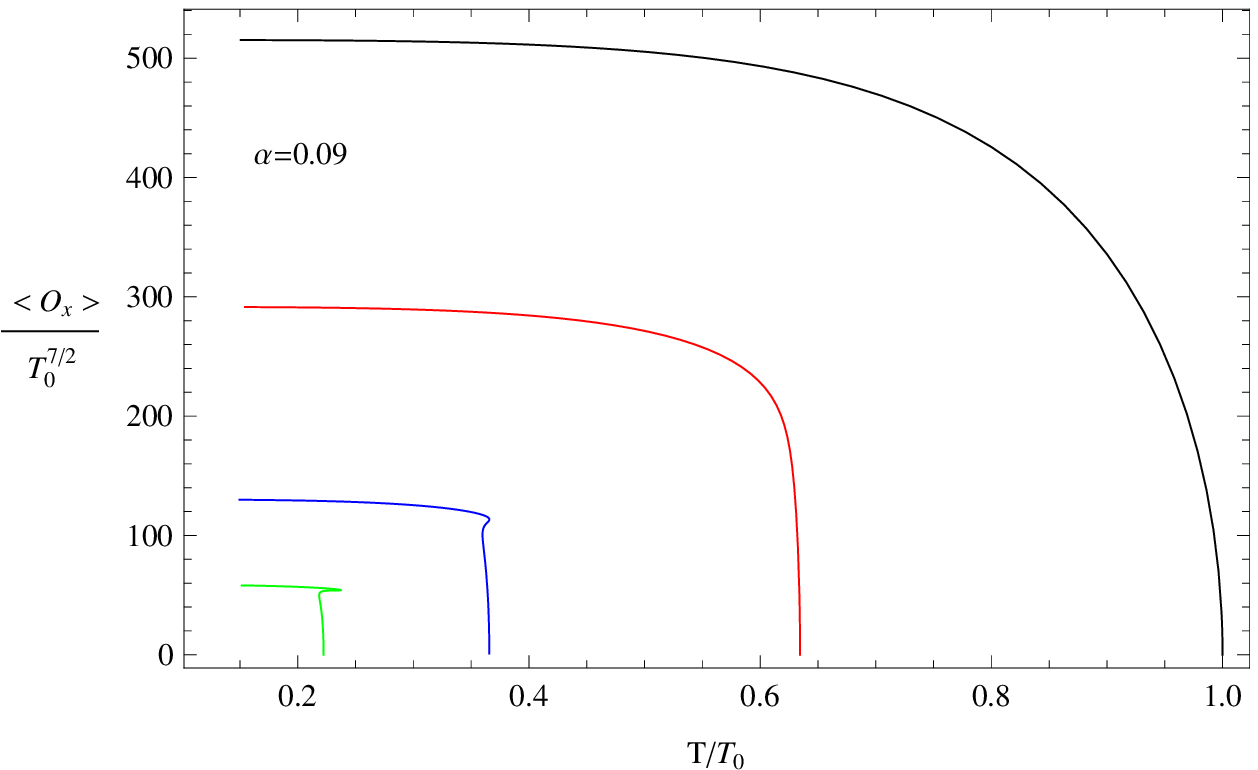}\hspace{0.2cm}%
\includegraphics[scale=0.425]{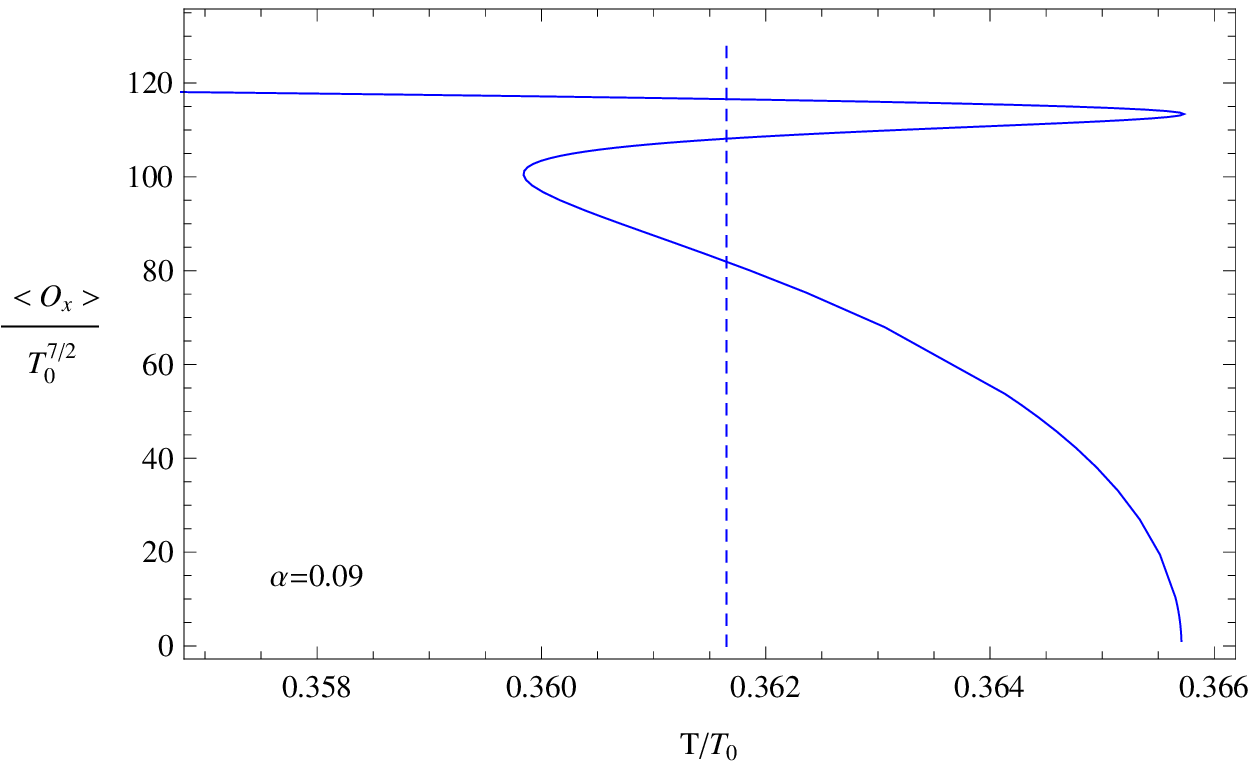}\hspace{0.2cm}%
\includegraphics[scale=0.425]{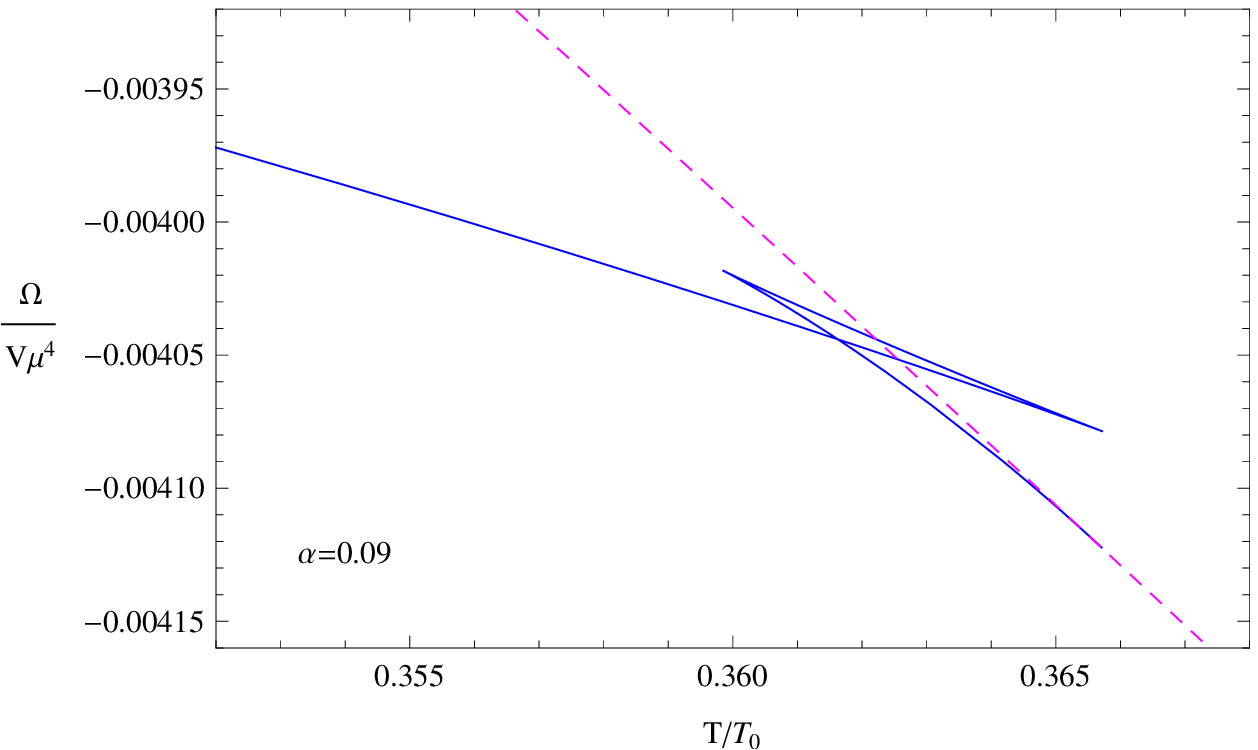}\\ \vspace{0.0cm}
\caption{\label{CondensateM125} (Color online) The condensate and
the grand potential as a function of the temperature with the fixed
mass of the vector field $m^{2}L^{2}_{eff}=5/4$ for different values
of the Gauss-Bonnet parameter $\alpha$. For the left three panels,
the four lines in each panel from top to bottom correspond to
increasing superfluid velocity, i.e., $\frac{S_{y}}{\mu}=0$ (black),
$0.50$ (red), $0.70$ (blue) and $0.80$ (green) respectively. For the
middle three panels, the line in each panel corresponds to the
superfluid velocity $\frac{S_{y}}{\mu}=0.70$ and a vertical line
represents the thermodynamically stable bound of the Cave of Winds.
For the right three panels, the two lines in each panel correspond
to the superfluid velocity $\frac{S_{y}}{\mu}=0.70$ (blue solid) and
the normal phase (magenta dotted) respectively.}
\end{figure}

In the case of the intermediate mass, such as
$m^{2}L^{2}_{eff}=5/4$, we plot the condensate and the grand
potential as a function of the temperature for different values of
$\alpha$ in Fig. \ref{CondensateM125}. For the vanishing or small
superfluid velocity, for example $\frac{S_{y}}{\mu}=0$ and $0.50$
with $\alpha=0.001$, the transition is second order and the
condensate approaches zero as $\langle
O_{x}\rangle\sim(T_{c}-T)^{\beta}$ with the mean field critical
exponent $\beta=1/2$ for all values of $\alpha$. However, when the
superfluid velocity improves beyond a special value, we observe that
$\langle O_{x}\rangle$ becomes multivalued and the Cave of Winds
appears, for example $\frac{S_{y}}{\mu}=0.70$ and $0.80$ with
$\alpha=0.001$. The thermodynamically favored region of the Cave of
Winds can be determined via its grand potential, just as in the
bottom two panels of the middle and rightmost columns in Fig.
\ref{CondensateM125}. Fixing $\frac{S_{y}}{\mu}=0.70$, from Fig.
\ref{CondensateM125} we find that the transition is second order in
the case of $\alpha=-0.19$ but the Cave of Winds appears in the case
of $\alpha=0.001$ or $\alpha=0.09$, which shows that the higher
curvature correction makes it easier for the emergence of the Cave
of Winds. As a matter of fact, the other choices of the intermediate
mass will not modify our result. This result seems to be interesting
since we can control the emergence of the Cave of Winds by using the
Gauss-Bonnet parameter. On the other hand, we also can see clearly
from Fig. \ref{CondensateM125} that the critical temperature
decreases with the increase of the Gauss-Bonnet parameter, which
indicates that the higher curvature correction hinders the phase
transition even the existence of the Cave of Winds.

\begin{figure}[ht]
\includegraphics[scale=0.6]{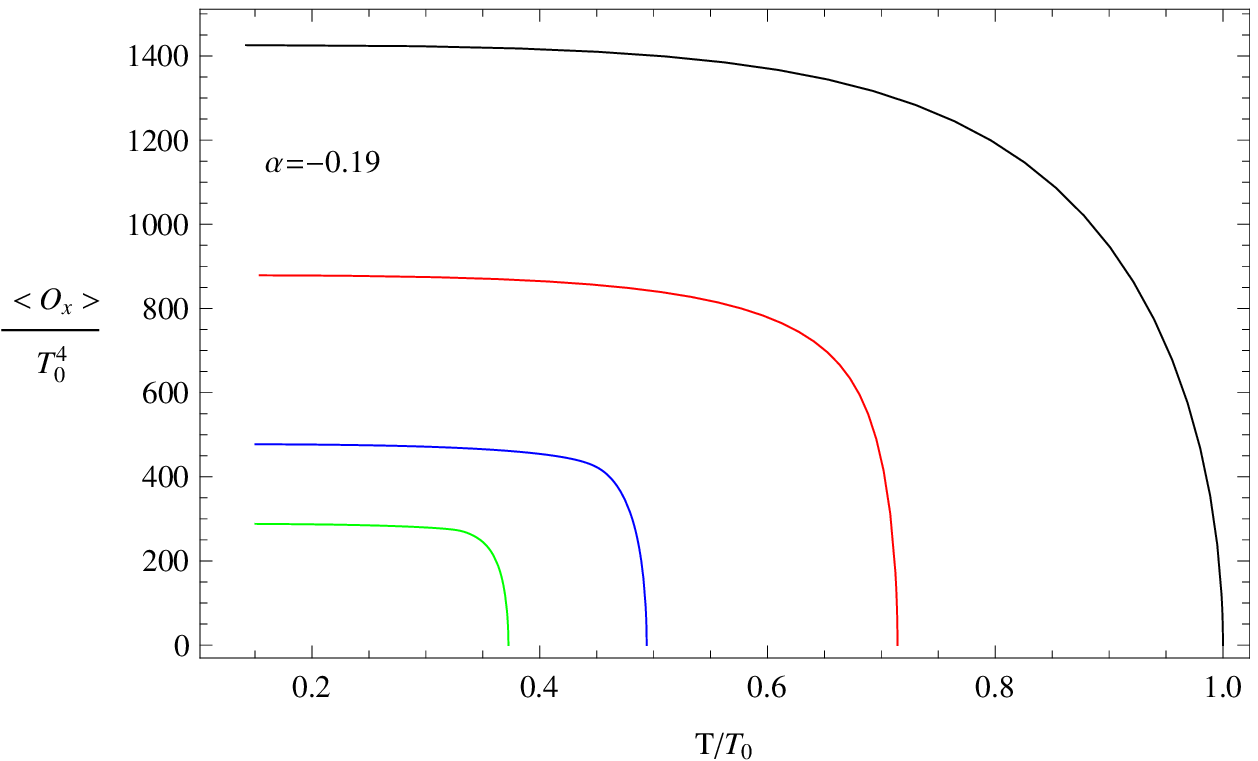}\hspace{0.2cm}%
\includegraphics[scale=0.6]{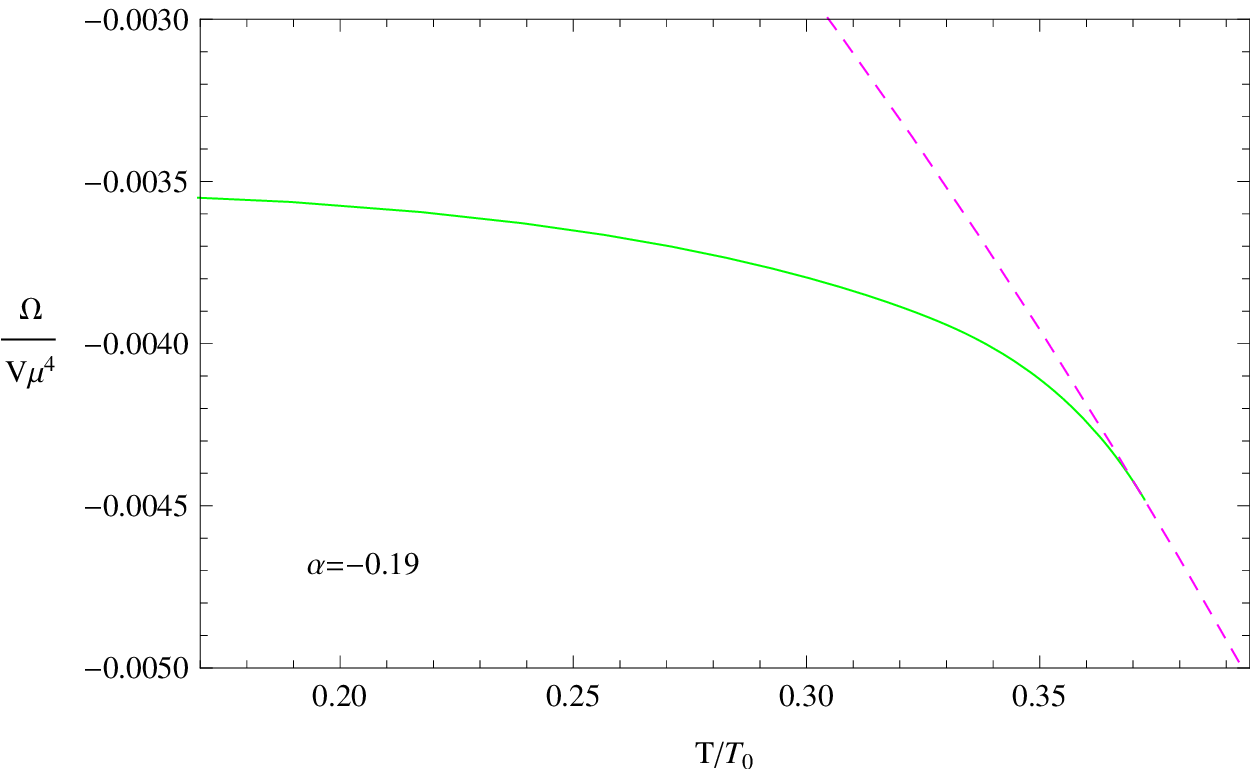}\\ \vspace{0.0cm}
\includegraphics[scale=0.6]{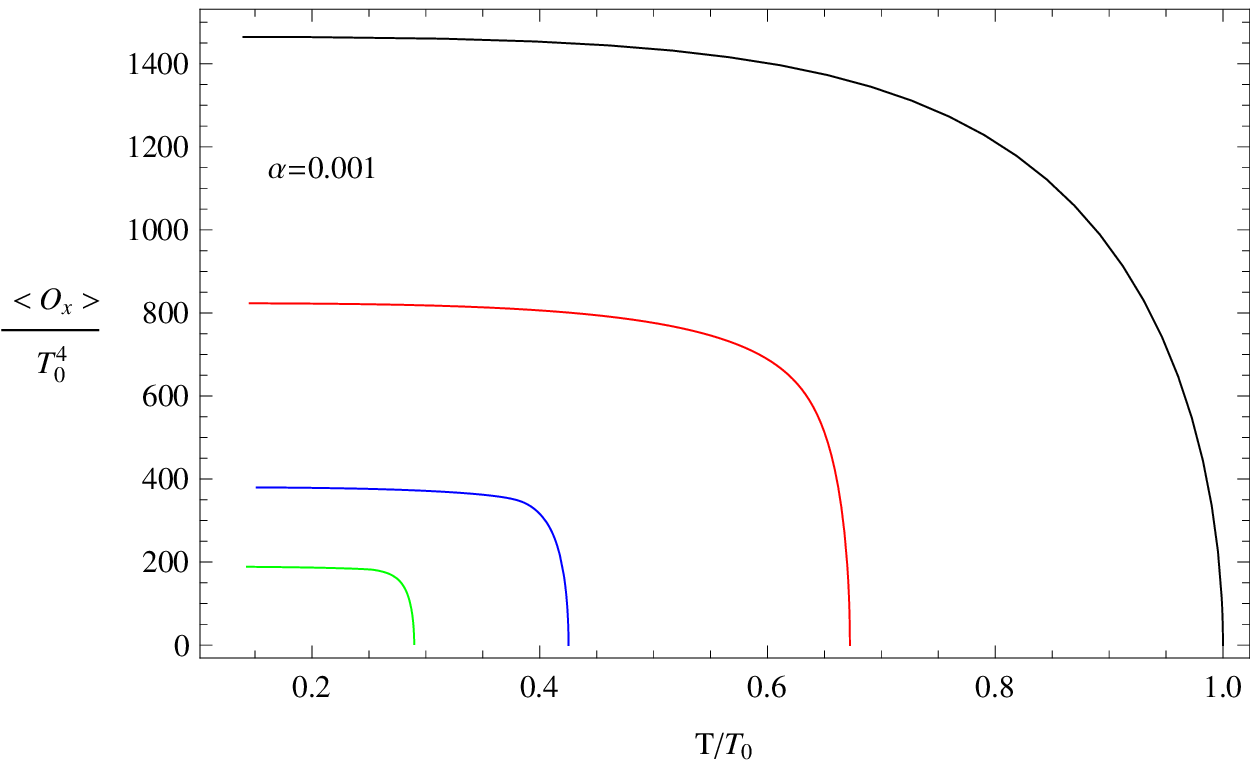}\hspace{0.2cm}%
\includegraphics[scale=0.6]{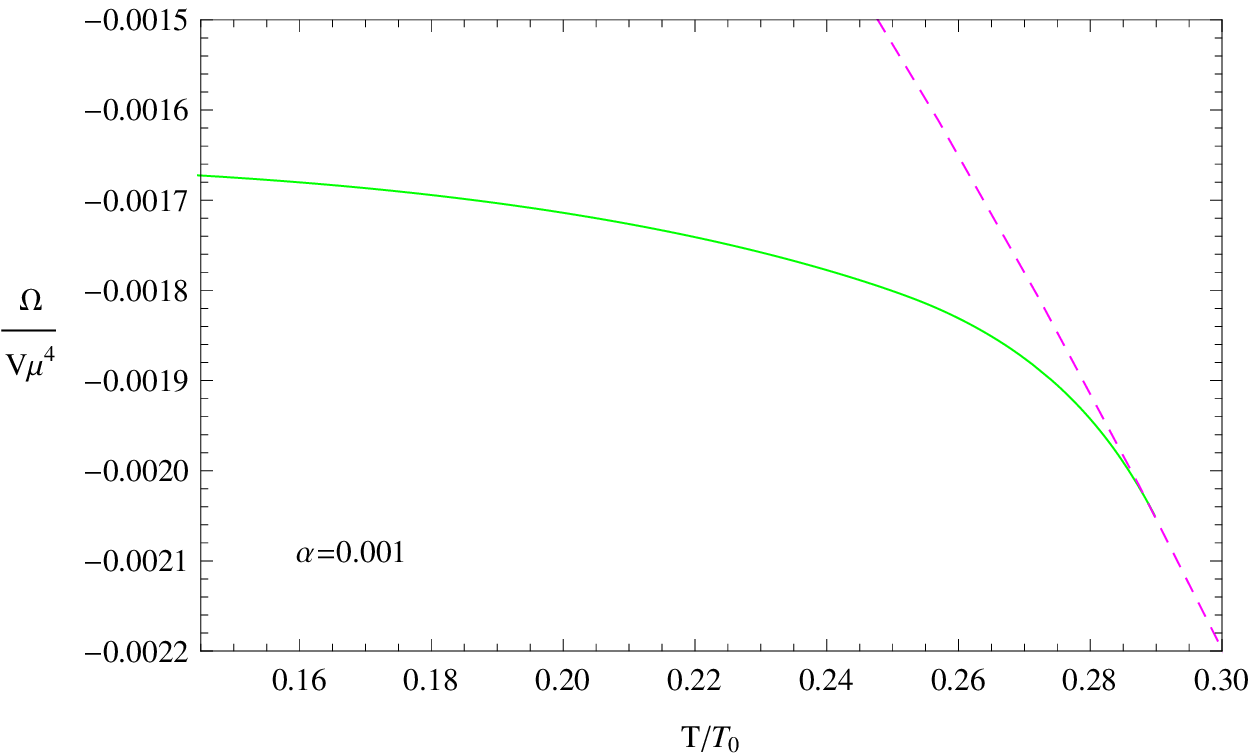}\\ \vspace{0.0cm}
\includegraphics[scale=0.6]{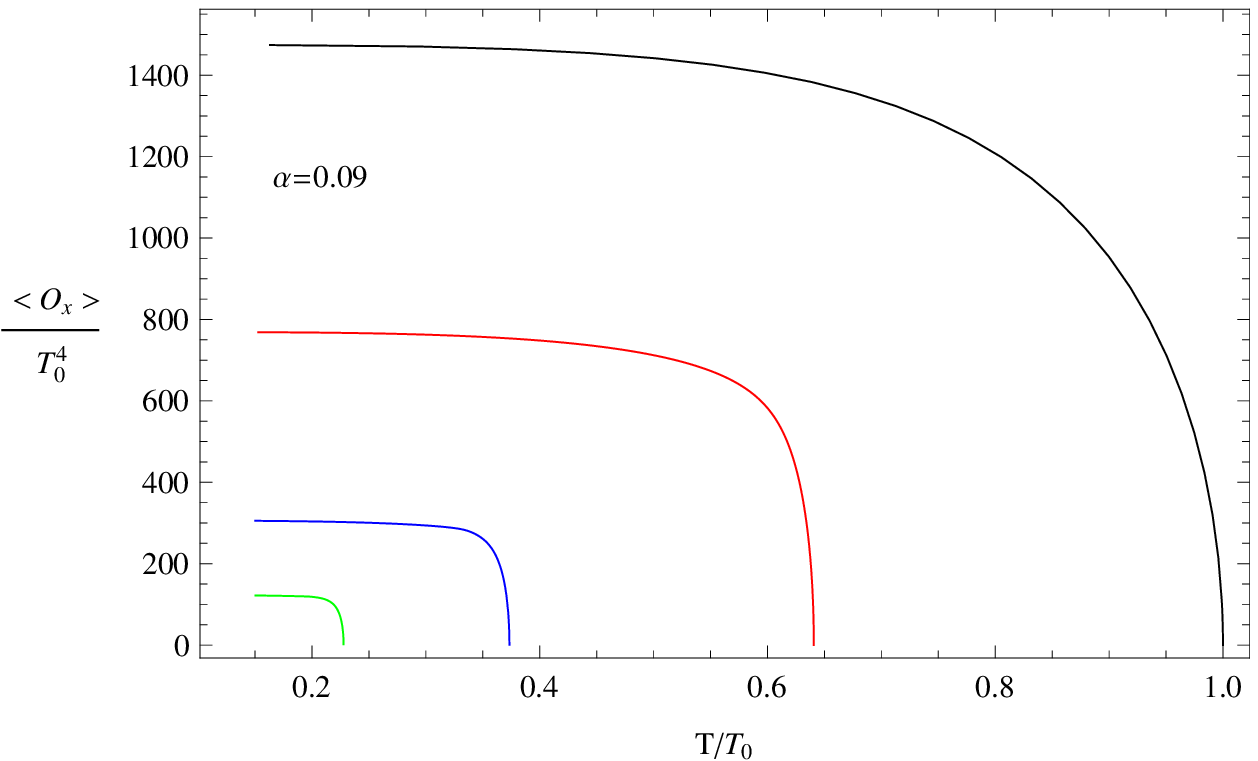}\hspace{0.2cm}%
\includegraphics[scale=0.6]{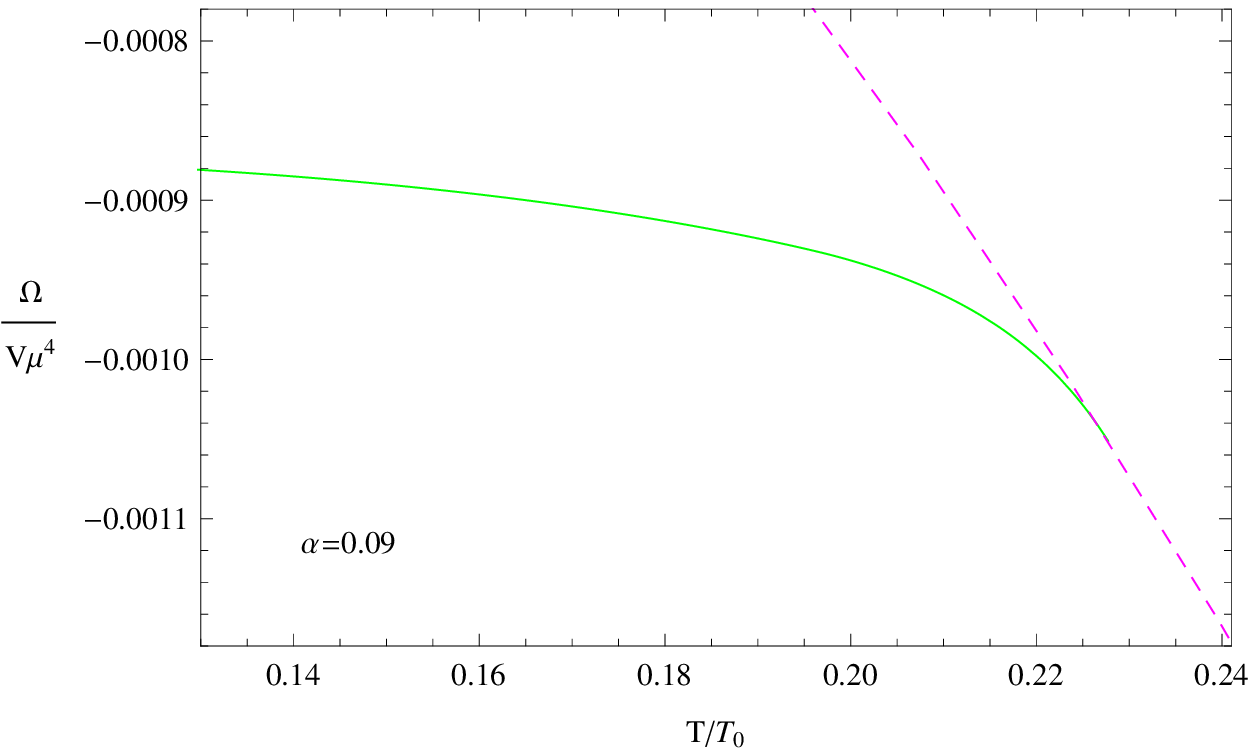}\\ \vspace{0.0cm}
\caption{\label{CondensateM300} (Color online) The condensate and
the grand potential as a function of the temperature with the fixed
mass of the vector field $m^{2}L^{2}_{eff}=3$ for different values
of the Gauss-Bonnet parameter $\alpha$. For the left three panels,
the four lines in each panel from top to bottom correspond to
increasing superfluid velocity, i.e., $\frac{S_{y}}{\mu}=0$ (black),
$0.5$ (red), $0.70$ (blue) and $0.80$ (green) respectively. For the
right three panels, the two lines in each panel correspond to the
superfluid velocity $\frac{S_{y}}{\mu}=0.80$ (green solid) and the
normal phase (magenta dotted) respectively.}
\end{figure}

Moving to the case of the sufficiently high mass, for example
$m^{2}L^{2}_{eff}=3$, we present in Fig. \ref{CondensateM300} the
condensate and the corresponding grand potential as a function of
the temperature for different values of $\alpha$. It should be noted
that, even in the rather high superfluid velocity
$\frac{S_{y}}{\mu}=0.80$, the phase transition of the system always
belongs to the second order and the Gauss-Bonnet parameter can not
change the order of phase transitions, which is consistent with the
corresponding grand potential in the three panels of the right
column in Fig. \ref{CondensateM300}. Since the effect of
$\frac{q^2A^2_y}{r^{2}}$ becomes relatively so weak that it can be
ignored for a high enough mass even in the Gauss-Bonnet gravity, our
results can be used to back up the findings obtained in Refs.
\cite{Arean2010,PWavSuperfluidA} that the system always suffers the
second-order phase transition in the case of a sufficiently high
mass.

\section{Supercurrents versus the superfluid velocity}

As the effective field theory of superconductors near the critical
temperature $T_{c}$, Ginzburg-Landau (GL) theory can give an
accurate description of such physical system and present various
significant quantities that can directly be compared with the
experimental results. Thus, we will study the relation between the
supercurrent and the superfluid velocity in this section and then
compare with GL theory.

\begin{figure}[ht]
\includegraphics[scale=0.425]{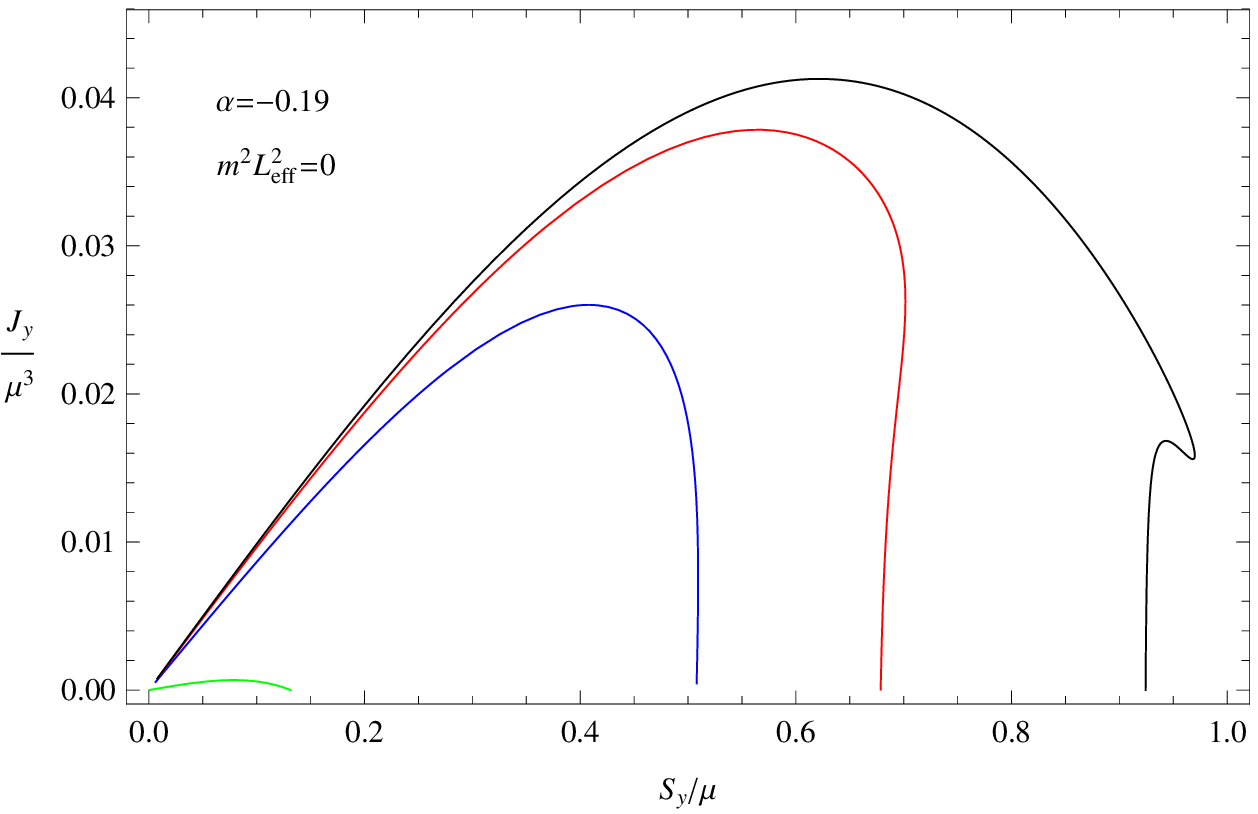}\hspace{0.2cm}%
\includegraphics[scale=0.425]{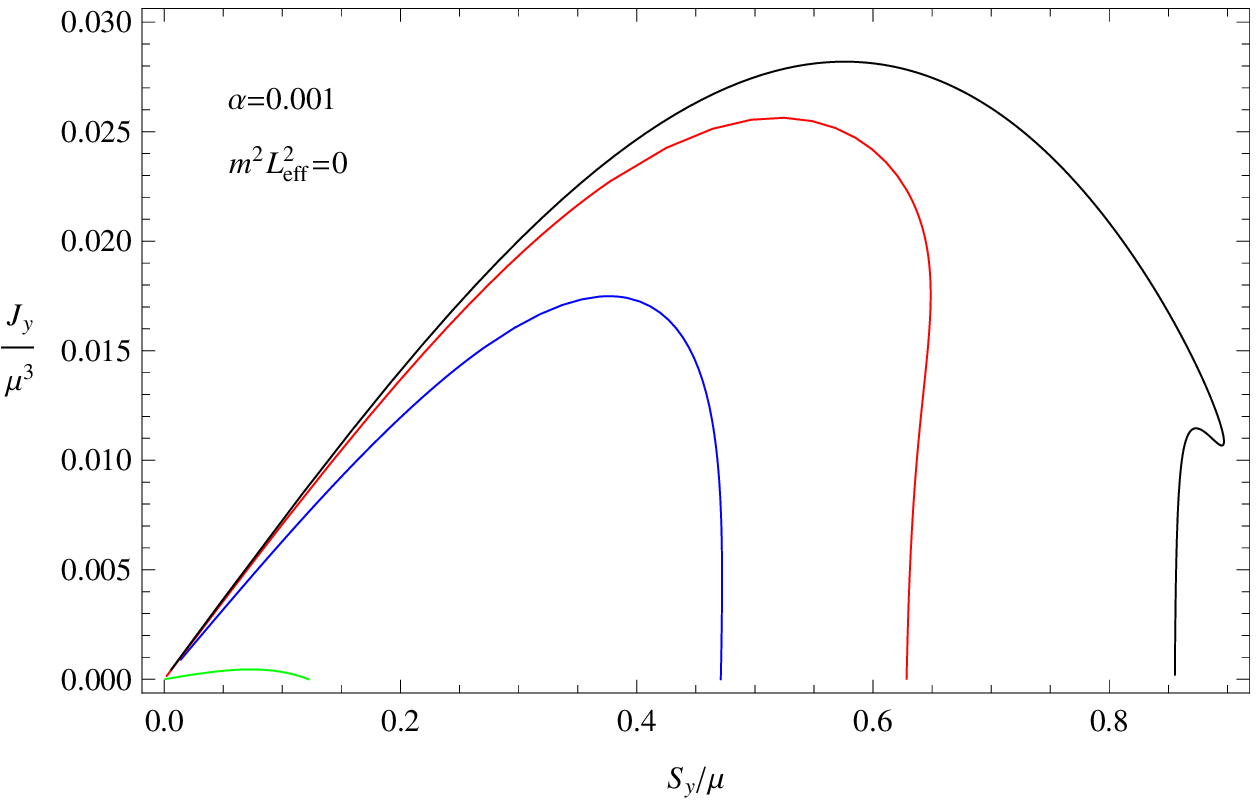}\hspace{0.2cm}%
\includegraphics[scale=0.425]{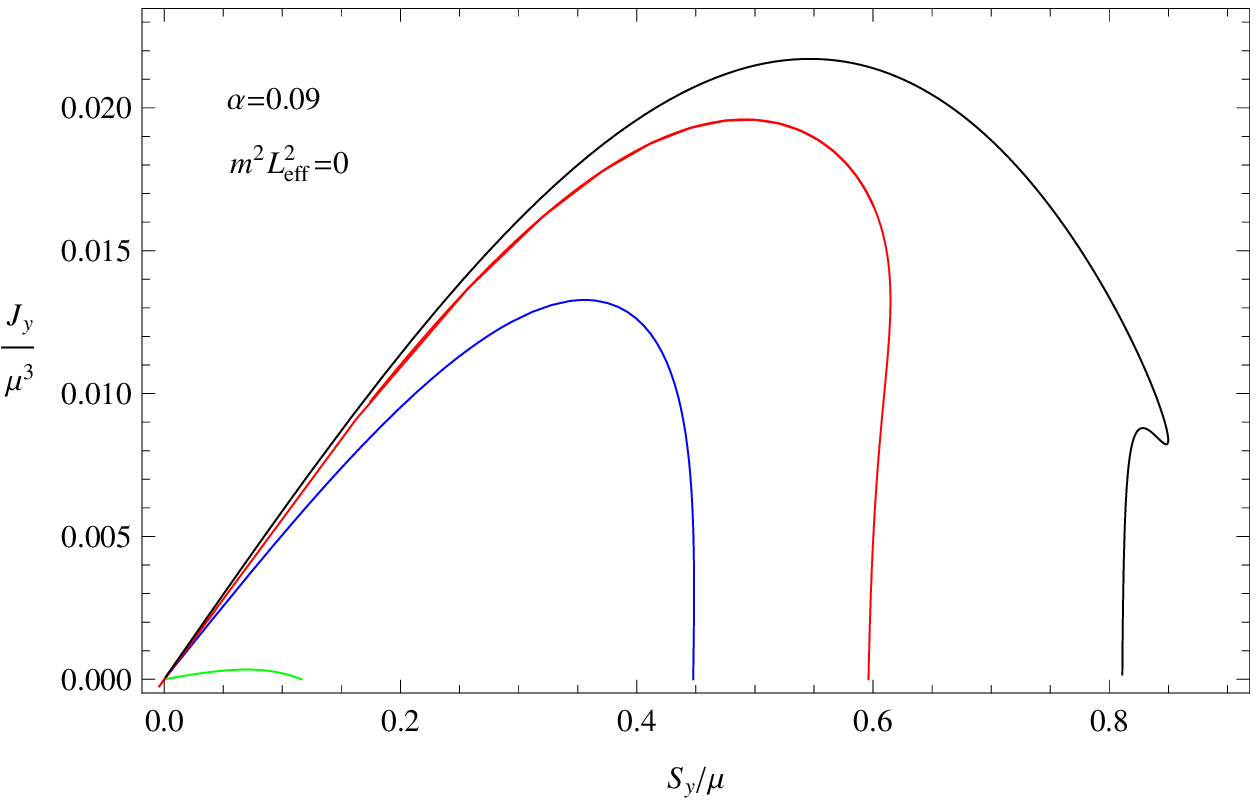}\\ \vspace{0.0cm}
\includegraphics[scale=0.425]{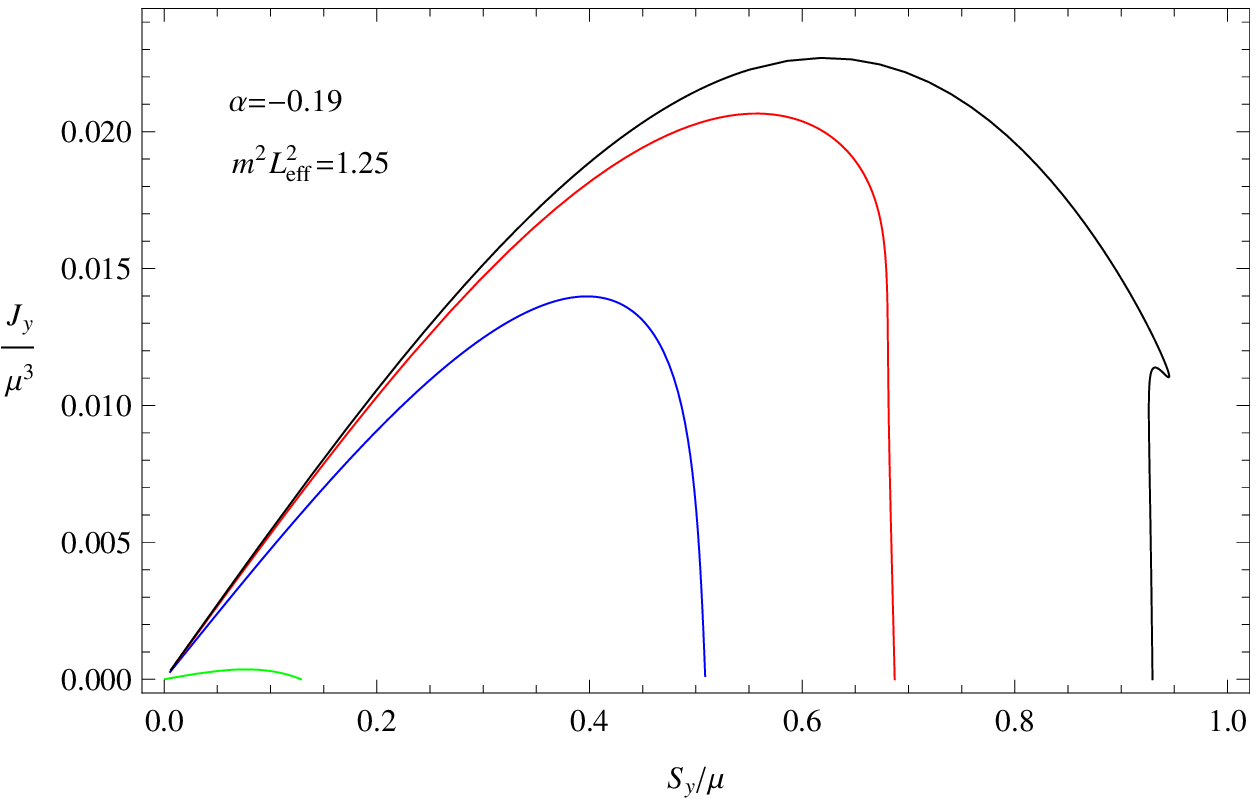}\hspace{0.2cm}%
\includegraphics[scale=0.425]{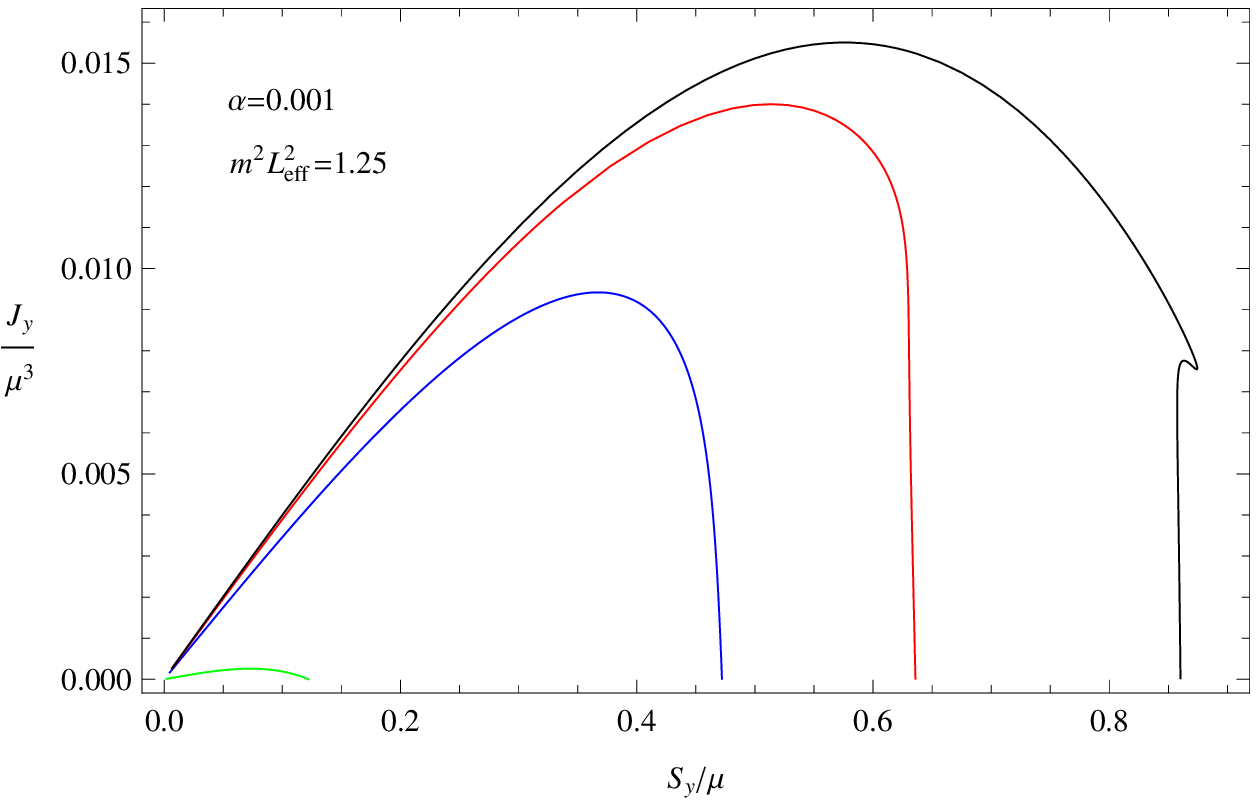}\hspace{0.2cm}%
\includegraphics[scale=0.425]{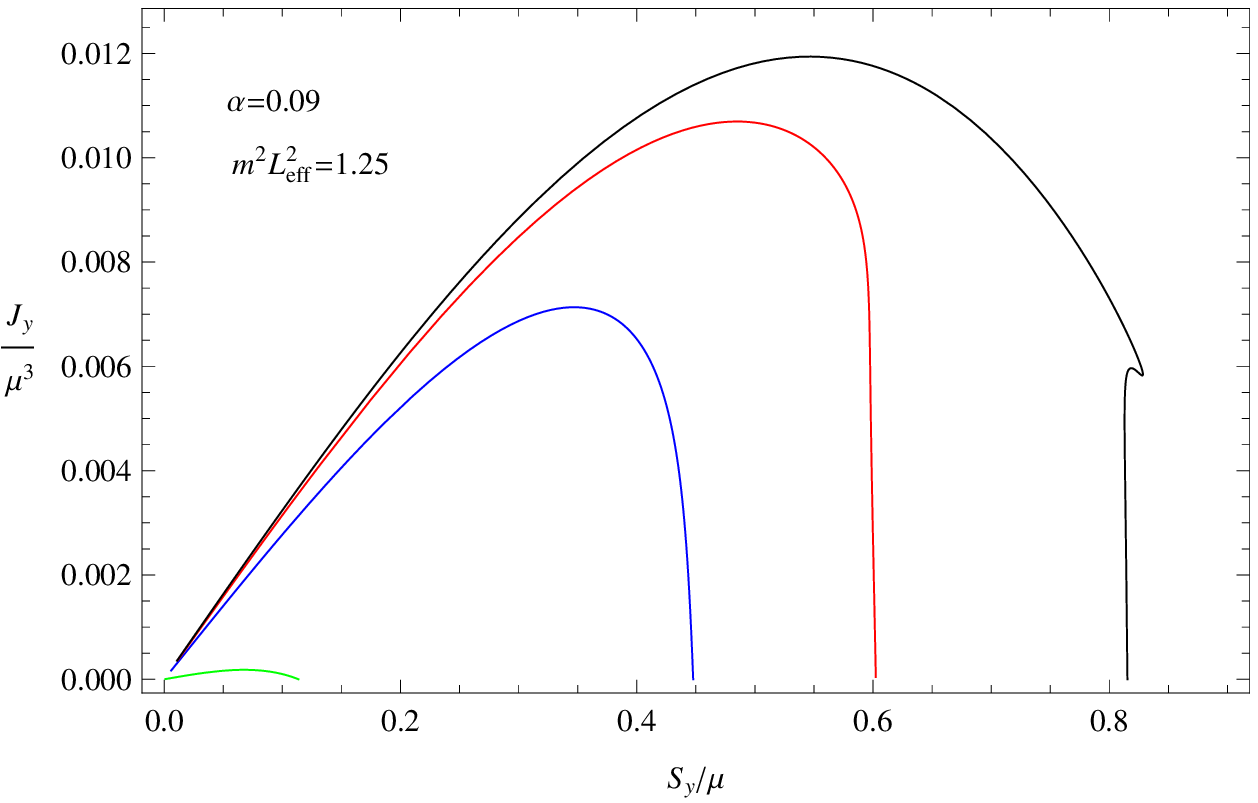}\\ \vspace{0.0cm}
\includegraphics[scale=0.425]{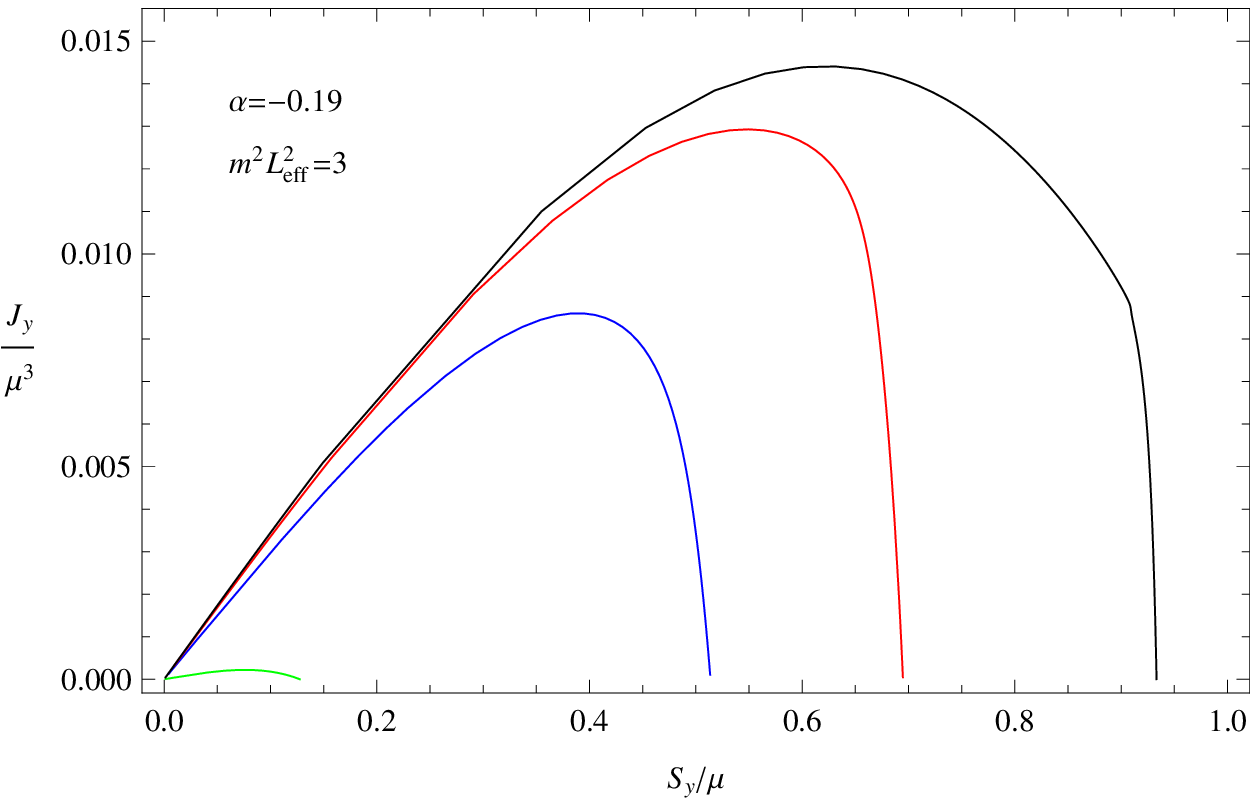}\hspace{0.2cm}%
\includegraphics[scale=0.425]{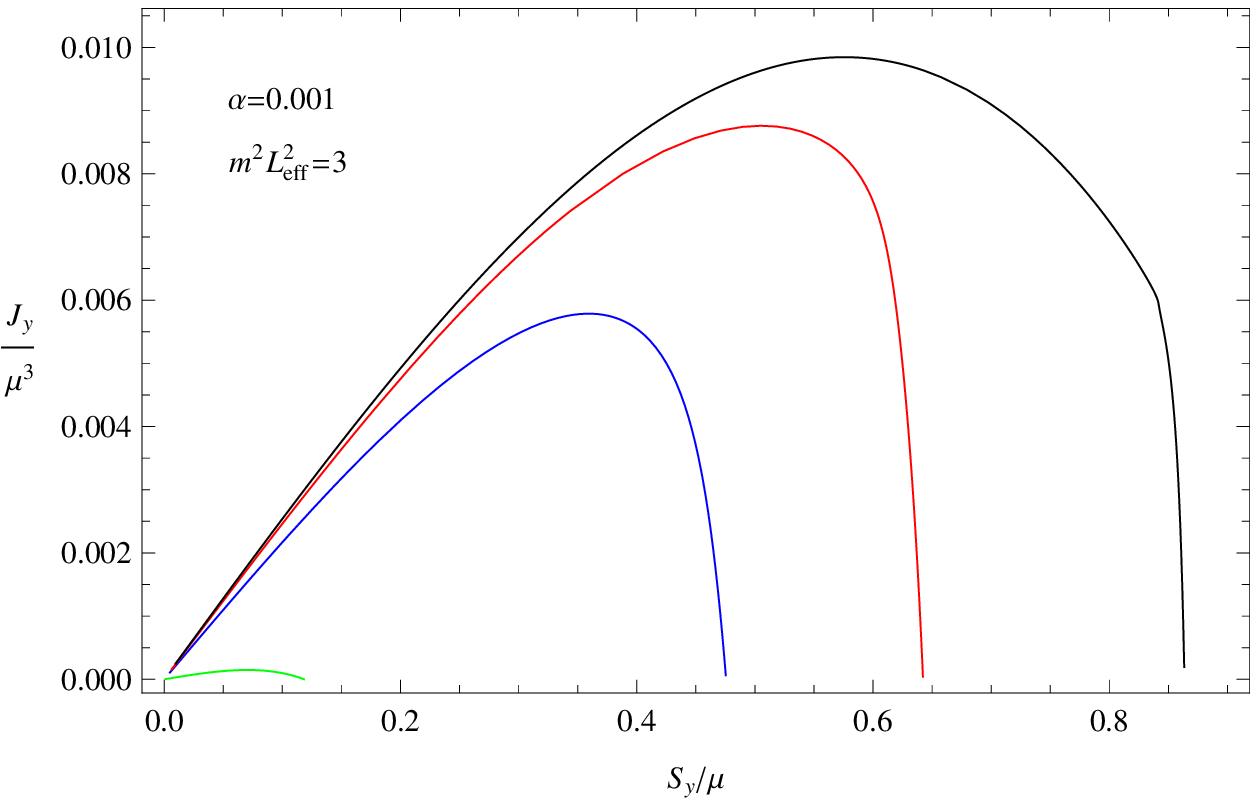}\hspace{0.2cm}%
\includegraphics[scale=0.425]{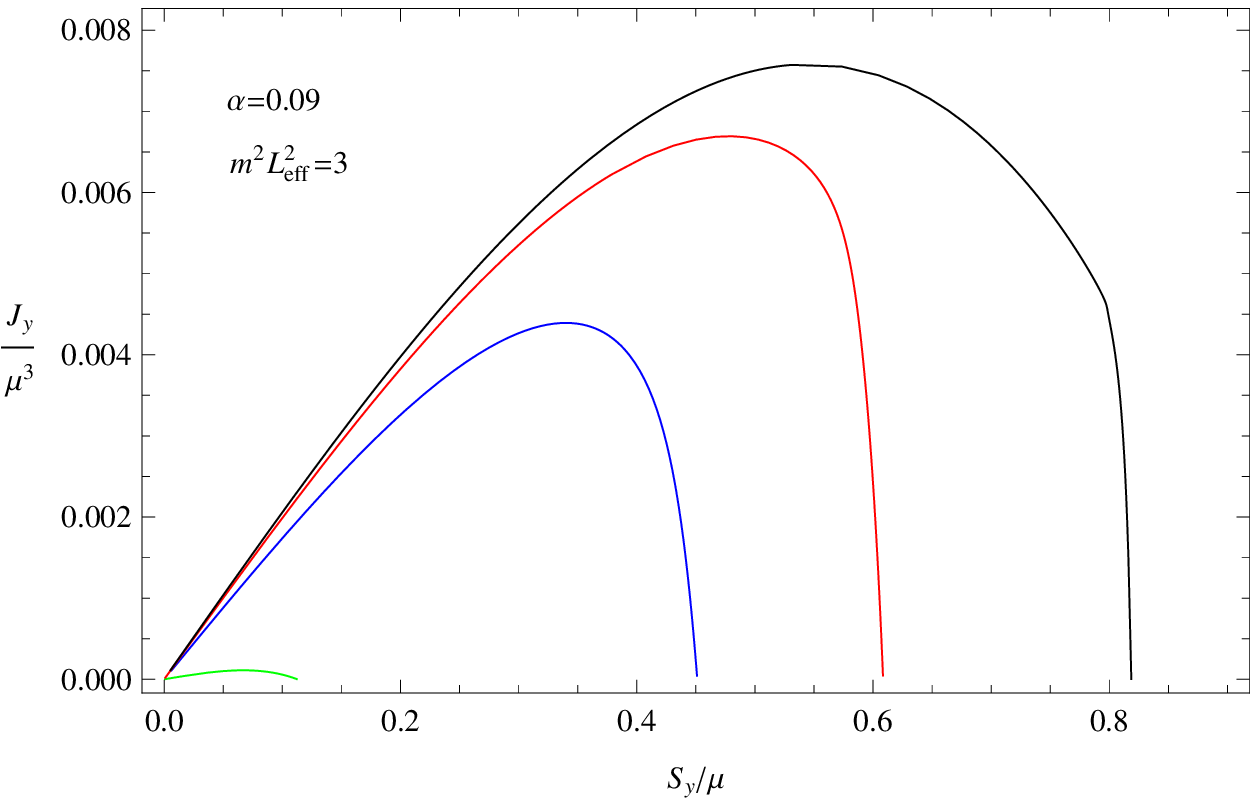}\\ \vspace{0.0cm}
\caption{\label{SupercurrentAlpha} (Color online) The supercurrent
versus superfluid velocity with the fixed mass of the vector field
for different values of the Gauss-Bonnet parameter $\alpha$. In each
panel, the four lines from right to left correspond to $T/T_{0}=0.2$
(black), $0.5$ (red), $0.7$ (blue) and $0.98$ (green) respectively.}
\end{figure}

In Fig. \ref{SupercurrentAlpha}, we plot the supercurrent as a
function of the superfluid velocity with the fixed mass of the
vector field $m^{2}L^{2}_{eff}=0$, $5/4$ and $3$ for different
values of the Gauss-Bonnet parameter $\alpha$. Due to the rich
phenomena in the phase transition, we have to discuss the behavior
of the supercurrent in an appropriate range of the vector field
mass. For small mass beyond the BF bound in the top three panels of
Fig. \ref{SupercurrentAlpha}, near the critical temperature, i.e.,
$T/T_{0}=0.98$, for all values of the Gauss-Bonnet parameter
$\alpha$ considered here, it is observed that the curves approximate
a parabola opening downward and the maximum value of the
supercurrent denoted by $\frac{J_{yMax}}{\mu^{3}}$ decreases with
the increasing $\alpha$. At the intersecting point of the larger
superfluid velocity with the abscissa axis, i.e.,
$\frac{S_{yMax}}{\mu}$, we find that the supercurrent
$\frac{J_{y}}{\mu^{3}}$ decreases smoothly to zero, which indicates
that the phase transition of the system belongs to the second order.
This result is in good agreement with the findings in the previous
section and also the same as that of GL theory. When the temperature
evidently deviates from the critical temperature, for example
$T/T_{0}=0.70$, the linear dependence of the supercurrent on the
superfluid velocity becomes more obvious until its maximum value
$\frac{J_{yMax}}{\mu^{3}}$, which agrees well with the one in the
thin superconducting films \cite{Tinkham}. When the superfluid
velocity increases and the temperature decreases to a certain value,
such as $T/T_{0}=0.50$ and $0.2$, the supercurrent versus the
superfluid velocity becomes double valued, which implies the latent
heat and the first-order phase transition in accordance with the
results in Fig. \ref{CondensateM0}. Especially, the larger the
Gauss-Bonnet parameter $\alpha$, the more easy it is for the curve
of the supercurrent to become double valued, which supports the
findings of Fig. \ref{TransSuperVelocity} and indicates that the
higher curvature correction makes it easier for the emergence of the
translating superfluid velocity from the second-order to the
first-order phase transition. For the intermediate mass in the
middle three panels of Fig. \ref{SupercurrentAlpha}, near the
critical temperature ($T/T_{0}=0.98$), the relation between the
supercurrent and the superfluid velocity is similar to the one in
the case of $m^{2}L^{2}_{eff}=0$, i.e., the curve of supercurrent
versus superfluid approximates a parabola opening downward and the
phase transition is of the second order. Decreasing the temperature
goes over a value, for example $T/T_{0}=0.20$, we see clearly that
the system first suffers a second-order transition and then a
first-order phase transition, but the critical point decreases as
the Gauss-Bonnet parameter increases, which is consistent with the
findings of Fig. \ref{CondensateM125} and means that the higher
curvature correction makes it easier for the emergence of the Cave
of Winds. For a high enough mass in the bottom three panels of Fig.
\ref{SupercurrentAlpha}, the holographic superfluid phase transition
always belongs to the second order at the temperatures considered
here, which is independent of the Gauss-Bonnet parameter $\alpha$
and in good agreement with the results in Fig. \ref{CondensateM300}.
On the other hand, we also find that the higher curvature correction
or larger mass, the lower the maximum value of the supercurrent
$\frac{J_{yMax}}{\mu^{3}}$, which agrees well with the results in
the previous section and suggests that the higher curvature
correction or larger mass makes it harder for the holographic
superfluid phase transition to be triggered.

\begin{figure}[ht]
\includegraphics[scale=0.425]{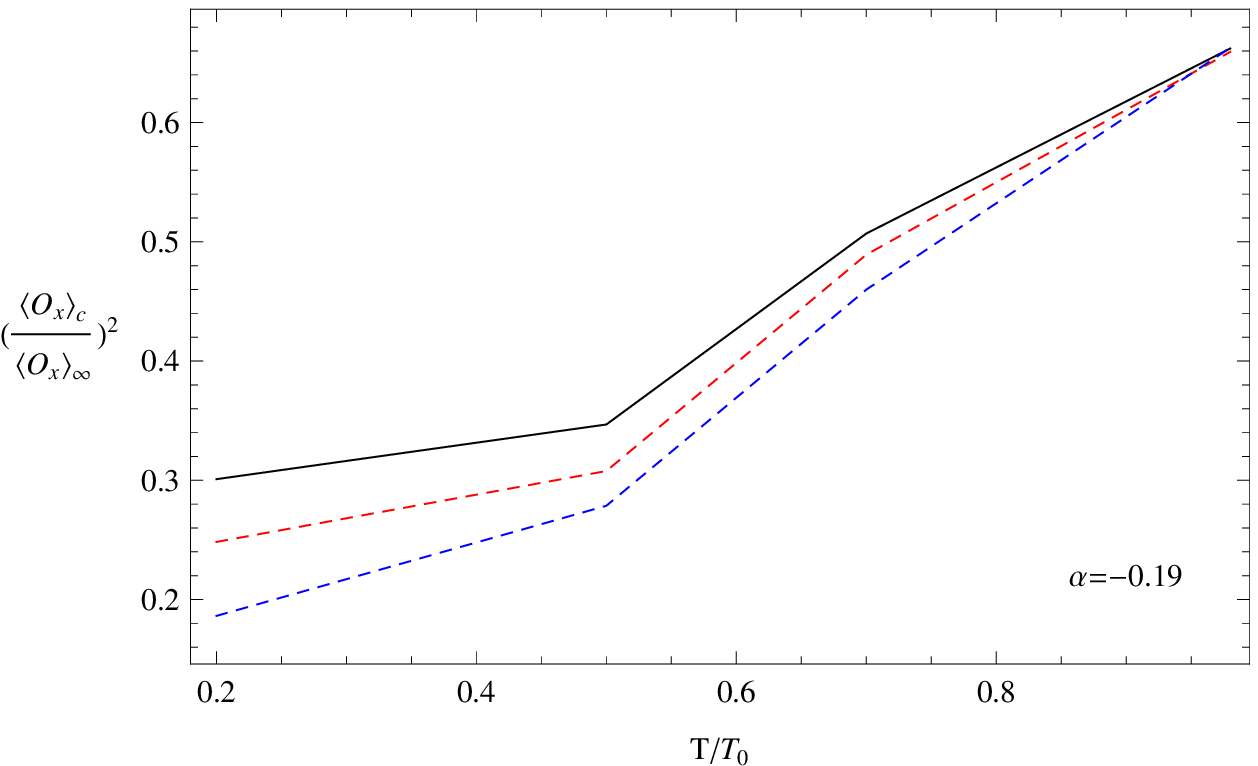}\hspace{0.2cm}%
\includegraphics[scale=0.425]{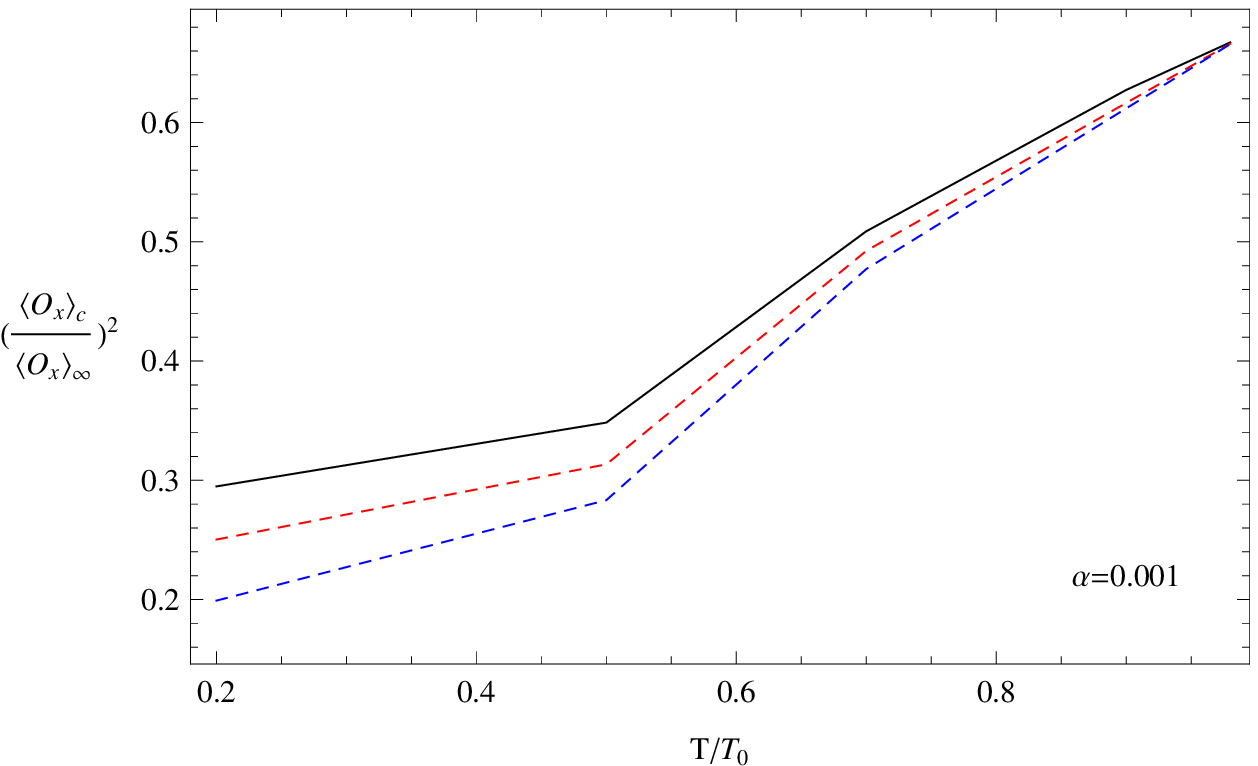}\hspace{0.2cm}%
\includegraphics[scale=0.425]{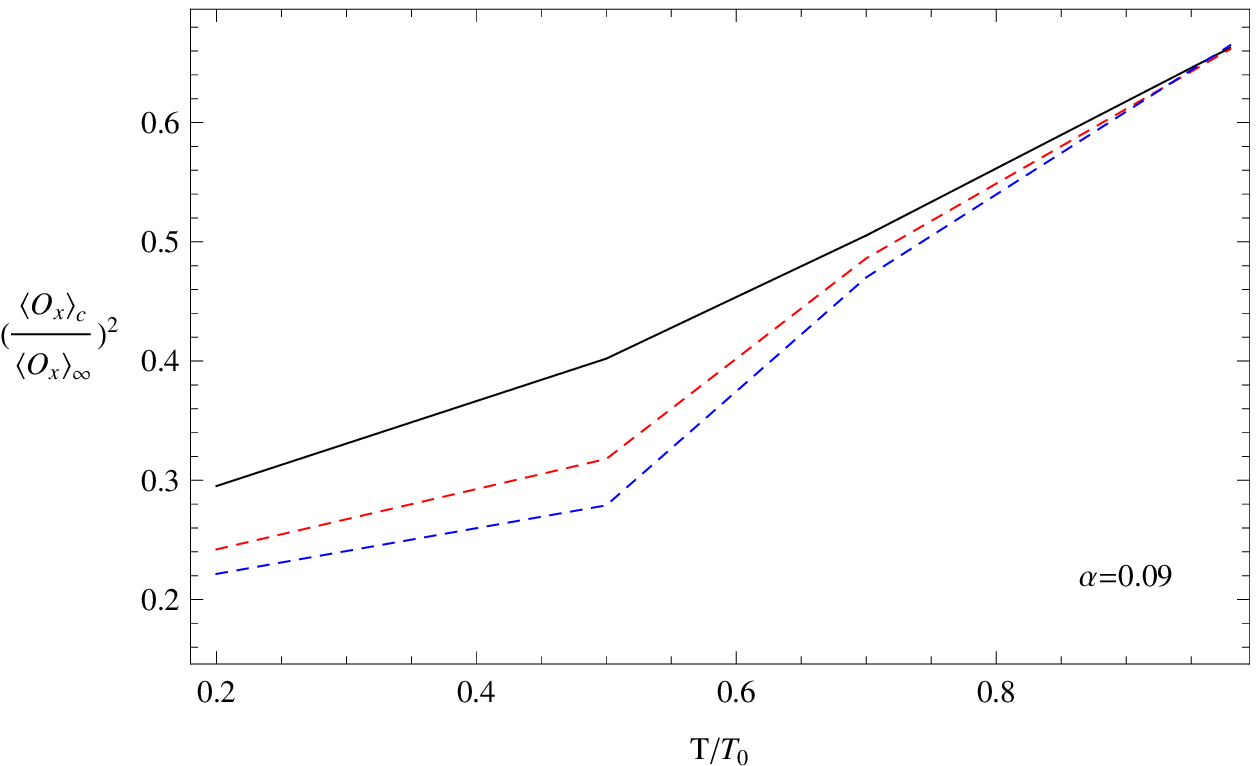}\\ \vspace{0.0cm}
\caption{\label{RatioAlpha} (Color online) The ratio $\left(\langle
O_x \rangle_c/{\langle O_x \rangle_\infty}\right)^2$ versus the
temperature with the fixed value of the Gauss-Bonnet parameter
$\alpha$ for different masses of the vector field. In each panel,
the three lines from top to bottom correspond to $m^2L_{eff}^2=0$
(black), $5/4$ (red) and $3$ (blue) respectively. }
\end{figure}

In order to check the reasonability of our holographic model
further, we will compare our results with another prediction of the
GL theory
\begin{eqnarray}\label{Ratio}
\left(\frac{\langle O_x \rangle_c}{\langle O_x
\rangle_\infty}\right)^2=\frac{2}{3},
 \end{eqnarray}
where $\langle O_x \rangle_\infty$ and $\langle O_x \rangle_c$ are
defined as the values of the condensate corresponding to the
vanishing superfluid velocity and the one with
$\frac{J_{yMax}}{\mu^{3}}$, respectively. Calculating the ratio
$\left(\langle O_x \rangle_c/{\langle O_x \rangle_\infty}\right)^2$
in our holographic p-wave superfluid model with Gauss-Bonnet
correction, we give the results with the fixed value of the
Gauss-Bonnet parameter $\alpha$ for different masses of the vector
field, i.e., $m^2L_{eff}^2=0$, $5/4$ and $3$ in Fig.
\ref{RatioAlpha}. From panels we observe that, near the critical
temperature, the ratio $\left(\langle O_x \rangle_c/{\langle O_x
\rangle_\infty}\right)^2$ is in very good agreement with GL theory
for all cases considered here, which is independent of the
Gauss-Bonnet parameter $\alpha$. However, the ratio deviates more
evidently from the predicted value $2/3$ when the temperature
decreases gradually from the critical temperature and the mass
increases from its BF bound. Interestingly, we observe that the
difference caused by the mass of the vector field is reduced when
the Gauss-Bonnet parameter becomes larger.

\section{conclusions}

In order to understand the influences of the $1/N$ or $1/\lambda$
($\lambda$ is the 't Hooft coupling) corrections on the holographic
dual models, we have constructed the holographic p-wave superfluid
model in the Gauss-Bonnet gravity via a Maxwell complex vector field
model and obtained the effect of the curvature correction on the
superfluid phase transition in the probe limit. We observed that,
regardless of the superfluid velocity and the vector field mass, the
critical temperature decreases as the Gauss-Bonnet parameter
increases, which indicates that the higher curvature correction
hinders the condensate of the vector field. It should be noted that
this conclusion still holds even in the case of the first-order
phase transition or the existence of the Cave of Winds. Considering
the rich phase structure of this system, we found that for the small
mass scale, the larger the Gauss-Bonnet parameter, the smaller the
translating superfluid velocity becomes, which shows that the higher
curvature correction makes it easier for the appearance of
translating point from the second-order transition to the
first-order one and implies that the Gauss-Bonnet parameter can
change the order of the phase transition in the holographic p-wave
superfluid system. In the case of the intermediate mass, we also
noted that the higher curvature correction makes it easier for the
emergence of the Cave of Winds. However, for the sufficiently high
mass, we showed that the phase transition of the system always
belongs to the second order and the Gauss-Bonnet parameter will not
affect the order of phase transitions. Furthermore, we investigated
the relation between the supercurrent and the superfluid velocity,
which is consistent with the findings obtained from the condensates
of the vector field in this model. In addition, we pointed out that,
near the critical temperature, the ratio $\left(\langle O_x
\rangle_c/{\langle O_x \rangle_\infty}\right)^2$ agrees well with
the GL theory, which is independent of the Gauss-Bonnet parameter.
However, the ratio deviates much more obviously as the temperature
decreases and the mass increases further. We found that the
difference in ratio caused by the mass of the vector field is
reduced when the Gauss-Bonnet parameter becomes larger. The
extension of this work to the fully backreacted spacetime would be
interesting since the backreaction provides richer physics in the
Maxwell complex vector field model \cite{CaiPWave-1,CaiPWave-2}. We
will leave it for further study.

\begin{acknowledgments}

We thank Dr. Jun-Wang Lu for his helpful discussions. This work was
supported by the National Natural Science Foundation of China under
Grant Nos. 11275066 and 11475061; Hunan Provincial Natural Science
Foundation of China under Grant No. 2016JJ1012.

\end{acknowledgments}

\end{document}